\documentclass[sigconf]{acmart}
\AtBeginDocument{%
  }

\usepackage{tikz}
\usepackage{amsmath}
\usepackage{multirow}
\usepackage{diagbox}
\usepackage{colortbl}
\usepackage{subcaption}

\setcopyright{acmlicensed}
\copyrightyear{2018}
\acmYear{}
\acmDOI{}

\acmSubmissionID{867}

\def\etal{\emph{et al.}}

\begin{document}
\definecolor{lightred}{rgb}{0.988,0.902,0.878}
\definecolor{lightgreen}{rgb}{0.973,1,0.765}


\title{Dark Distillation: Backdooring Distilled Datasets without Accessing Raw Data}

\author{Ziyuan Yang}
\authornote{This work was conducted during the visiting tenure of Ziyuan Yang at the Agency for Science, Technology and Research (A*STAR).}
\email{cziyuanyang@gmail.com}
\affiliation{
\institution{Agency for Science, Technology and Research (A*STAR)}
\country{Singapore}
}

\author{Ming Yan}
\email{Yan\_Ming@cfar.a-star.edu.sg}
\affiliation{
\institution{Agency for Science, Technology and Research~(A*STAR)}
\country{Singapore}
}

\author{Yi Zhang}
\authornote{Corresponding Author}
\email{yzhang@scu.edu.cn}
\affiliation{
\institution{Sichuan University}
  \country{China}
}

\author{Joey Tianyi Zhou}
\authornotemark[2]
\email{Joey\_Zhou@cfar.a-star.edu.sg}
\affiliation{
\institution{Agency for Science, Technology and Research~(A*STAR)}
\country{Singapore}
}


\begin{abstract}

Dataset distillation (DD) is a powerful technique that enhances training efficiency and reduces transmission bandwidth by condensing large datasets into smaller, synthetic ones. It enables models to achieve performance comparable to those trained on the raw full dataset and has become a widely adopted method for data sharing. However, security concerns in DD remain underexplored. Existing studies typically assume that malicious behavior originates from dataset owners during the initial distillation process, where backdoors are injected into raw datasets. In contrast, this work is the first to address a more realistic and concerning threat: attackers may intercept the dataset distribution process, inject backdoors into the distilled datasets, and redistribute them to users. While distilled datasets were previously considered resistant to backdoor attacks, we demonstrate that they remain vulnerable to such attacks. Furthermore, we show that attackers do not even require access to any raw data to inject the backdoors successfully. Specifically, our approach reconstructs conceptual archetypes for each class from the model trained on the distilled dataset. Backdoors are then injected into these archetypes to update the distilled dataset. Moreover, we ensure the updated dataset not only retains the backdoor but also preserves the original optimization trajectory, thus maintaining the knowledge of the raw dataset. To achieve this, a hybrid loss is designed to integrate backdoor information along the benign optimization trajectory, ensuring that previously learned information is not forgotten. Extensive experiments demonstrate that distilled datasets are highly vulnerable to backdoor attacks, with risks pervasive across various raw datasets, distillation methods, and downstream training strategies. Moreover, our attack method is highly efficient and lightweight, capable of synthesizing a malicious distilled dataset in under one minute in certain cases~\footnote{The code will be made publicly available to ensure reproducibility.}.
\end{abstract}





\maketitle

\section{Introduction}
Deep learning (DL) has achieved remarkable success in recent years, driven by advancements in computational resources and large-scale datasets~\cite{lei2023comprehensive}. With the rise of large language models, such as GPT-3, which has 175 billion parameters and was trained on 45 terabytes of text data using thousands of GPUs for a month~\cite{brown2020language}, the demand for computational power and data has reached unprecedented levels. However, the exponential growth of data has created a significant imbalance with computational capacity, posing challenges to training efficiency and costs~\cite{lecun2015deep}.

Dataset distillation (DD) has recently emerged as a promising solution to the challenges posed by large-scale datasets and their computational demands~\cite{dudiversity}. By synthesizing smaller datasets that retain the essential information of the raw data, DD enables efficient training while significantly reducing storage and computational costs, with minimal impact on model performance~\cite{sun2024diversity}. With advantages such as lower storage, training, and energy costs, DD is expected to become a widely adopted method for data sharing, playing a pivotal role in many machine learning applications~\cite{yu2024teddy}. 

Most existing DD methods focus solely on preserving the information of the raw dataset, often overlooking security issues. While these issues have recently garnered some attention from researchers, the number of related studies remains limited. For example, Liu~\etal~\cite{liu2023backdoor} proposed DoorPing, a learnable trigger that is iteratively updated during the distillation procedure. Similarly, Chung~\etal~\cite{chung2024rethinking} introduced a standard optimization framework to learn triggers for DD. 

\begin{figure*}[!h]
  \centering
  \includegraphics[width=\linewidth]{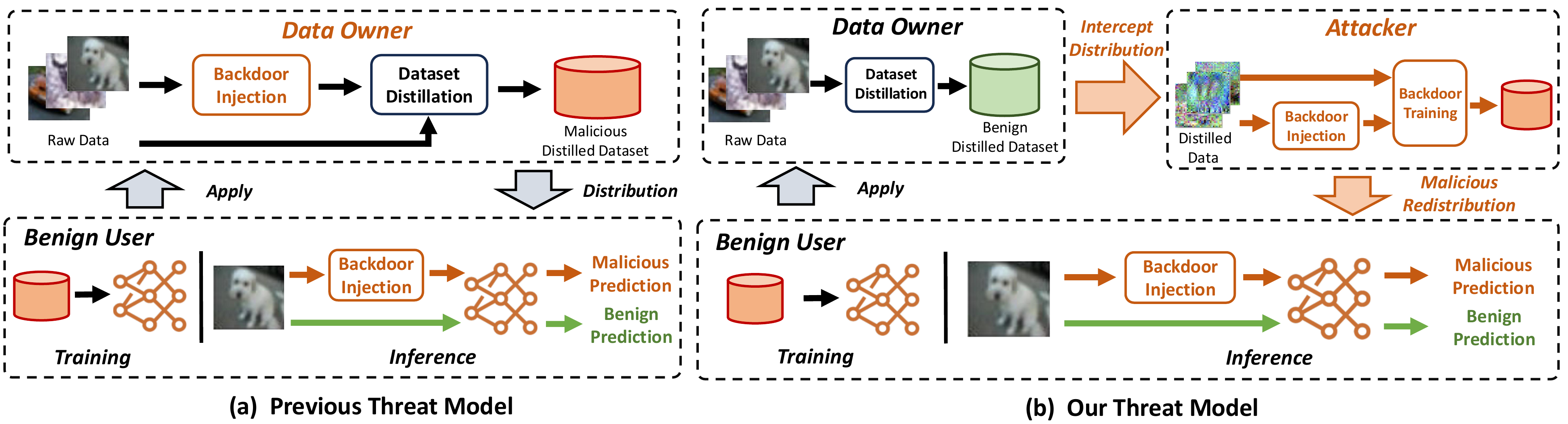}
  \vspace{-20pt}
  \caption{Illustration of the threat models. (a) Previous works assume that the data owner may be malicious and inject backdoors into the distilled dataset before distributing it to users. (b) In contrast, our threat model is more practical. We assume the data owner is benign. However, third parties, such as hackers or malicious users, may act maliciously. They could attack the system by hijacking the dataset distribution, injecting backdoors into the distilled dataset, and redistributing it to users.}
  \vspace{-10pt}
  \label{fig:toy_example}
\end{figure*}

However, the threat models of these methods assume that the dataset owner intentionally injects backdoors during the distillation process.
In practice, dataset owners are unlikely to compromise their own data by injecting backdoors. Instead, a more plausible threat arises from third-party adversaries. For instance, during dataset distribution, attackers may intercept access to a benign distilled dataset, inject backdoors, and redistribute the compromised version to unsuspecting users, enabling malicious activities.
Additionally, distilled datasets are often considered privacy-preserving~\cite{dong2022privacy}, secure~\cite{liu2023backdoor}, and highly compact, making them suitable for storage on various Internet-of-Things (IoT) devices or clients in distributed learning paradigm~\cite{zhu2021data,xu2024privacy}. This widespread deployment increases the risk of unauthorized access, facilitating manipulation of the dataset by attackers. Once compromised, the backdoored distilled dataset can be redistributed to other users, thereby amplifying the threat.
To highlight the distinction between previous threat models and ours, we provide an illustrative example in Figure~\ref{fig:toy_example}.


In this work, we consider a practical threat model where malicious behavior originates from third parties during data sharing. Specifically, we attempt to directly inject backdoors into the distilled dataset while ensuring that the malicious behavior can still be triggered by real images. This represents a particularly difficult attack assumption for attackers, as it relies on the premise that the malicious third party does not have access to any raw data. Moreover, the significant gap between synthetic and real images presents an additional challenge.

To address these challenges and evaluate the vulnerabilities of distilled datasets, we propose a novel and the first backdoor attack method specifically designed for this threat model. Under our strict assumption, the attacker has no access to raw data. However, the fundamental paradigm of DD involves synthesizing small-scale datasets that retain the knowledge of the raw dataset. This implies that the distilled dataset inherently encapsulates the knowledge of the raw data. While it is almost impossible to reconstruct visually similar images to the raw data without any prior knowledge, the inherent properties of DL enable us to focus on the deep feature space. We only need to ensure that the reconstructed images in the latent feature space share a similar distribution with real images, which allows the trigger to be effectively activated in this space. Leveraging this paradigm, we aim to reconstruct conceptual archetypes for each class, derived from the knowledge embedded in the model trained on the benign distilled dataset, to serve as the foundation of our attack.

Next, we inject backdoors into these conceptual archetypes while ensuring that the modified distilled dataset retains the knowledge of the raw dataset. To achieve this, we propose a hybrid loss function that injects backdoor information into the malicious distilled dataset while preserving the original optimization trajectory. This approach bridges the gap between the distilled dataset and real images, ensuring that the backdoor can be reliably activated by real images while minimizing performance degradation for benign images.

Notably, our method directly injects backdoors into the distilled dataset without requiring prior knowledge of the DD method, raw data, or the downstream model. Extensive experiments demonstrate that our approach can successfully compromise the security of distilled datasets, regardless of the DD method, downstream model architecture, or training strategy. This finding challenges the prevailing belief that distilled datasets are inherently secure~\cite{liu2023backdoor} and reveals significant security vulnerabilities. Additionally, our attack method is highly lightweight, capable of synthesizing malicious distilled datasets within one minute in certain scenarios.
The main contributions can be summarized as follows:

\begin{itemize}
    \item We investigate a novel threat model for DD, where backdoors are directly injected into distilled datasets without requiring any raw data access. To the best of our knowledge, this is the first study to explore this threat in DD.
    \item We propose the first backdoor injection method for distilled datasets
    that reconstructs conceptual archetypes and injects backdoors while preserving the knowledge of the raw dataset.
    \item We design a hybrid loss to ensure the backdoor injection aligns with the original optimization trajectory, maintaining backdoor activation in real images while minimizing performance degradation on benign tasks.
    \item Extensive experiments across diverse datasets, DD methods, networks, and training strategies validate the generalizability of our method and expose DD vulnerabilities. Moreover, our attack is highly efficient, synthesizing malicious distilled datasets in under a minute in certain cases.
\end{itemize}

\section{Related Works}
\noindent \textbf{Dataset Distillation.} DD aims to condense the richness of large-scale datasets into compact small datasets that effectively preserve training performance~\cite{yu2023dataset}.
Coreset selection~\cite{du2024sequential} is an early-stage research in data-efficient learning. Most methods rely on heuristics to select representatives. 
Unlike this paradigm, DD~\cite{wang2018dataset} aims to learn how to synthesize a tiny dataset that trains models to perform comparably to those trained on the complete dataset. Wang~\textit{et al.}~\cite{wang2018dataset} first proposed a bi-level meta-learning approach, which optimizes a synthetic dataset so that neural networks trained on it achieve the lowest loss on the raw dataset.

Following this research, many researchers have focused on reducing the computational cost of the inner loop by introducing closed-form solutions, such as kernel ridge regression~\cite{loo2022efficient,chenprovable,xu2023kernel}. 
Zhao~\textit{et al.}~\cite{zhao2021dataset} proposed an approach that makes parameters trained on condensed data approximate the target parameters, formulating a gradient matching objective that simplifies the DD process from a parameter perspective. In~\cite{zhao2021datasetdsa}, the authors enhanced the process by incorporating Differentiable Siamese Augmentation~(DSA), which enables effective data augmentation on synthetic data and results in the distillation of more informative images. Additionally, Du~\textit{et al.}~\cite{du2024sequential} proposed a sequential DD method to extract the high-level features learned by the DNN in later epochs. 
By combining meta-learning and parameter matching, Cazenavette~\textit{et al.}~\cite{cazenavette2022dataset} proposed Matching Training Trajectories~(MTT) and achieved satisfactory performance. Besides, a recent work, TESLA~\cite{cui2023scaling}, reduced GPU memory consumption and can be viewed as a memory-efficient version of MTT. 

\noindent \textbf{Backdoor Attack.} Backdoor attacks introduce malicious behavior into the model without degrading its performance on the original task by poisoning the dataset. Gu~\textit{et al.}~\cite{gu2019badnets} introduced the backdoor threat in DL with BadNets, which injects visible triggers into randomly selected training samples and mislabels them as a specified target class. To enhance attack stealthiness, Chen~\textit{et al.}~\cite{chen2017targeted} proposed a blended strategy to make poisoned images indistinguishable from benign ones, improving their ability to evade human inspection. Furthermore, subsequent works explored stealthier attacks: WaNet~\cite{nguyen2020wanet} used image warping; ISSBA~\cite{li2021invisible} employed deep steganography; Feng~\textit{et al.}~\cite{feng2022fiba} and Wang~\textit{et al.}~\cite{wang2022invisible} embedded triggers in the frequency domain; Yang~\textit{et al.}~\cite{yang2024inject} injected the trigger into the measurement domain; and Color Backdoor~\cite{jiang2023color} utilized uniform color space shifts as triggers. 

Although existing works have demonstrated the vulnerability of deep networks to backdoor attacks, the exploration of such vulnerabilities in the context of DD remains limited. Only a few studies have evaluated the security risks associated with DD~\cite{liu2023backdoor,chung2024rethinking}. This highlights the urgent need for a deeper investigation into the potential threats and vulnerabilities specific to DD. 


\section{Threat Model}

In previous works~\cite{liu2023backdoor,chung2024rethinking}, the threat model assumes all users are benign, the data owner is malicious, and the attack method has access to the raw data and knowledge of the specific DD method used. These are highly restrictive and unrealistic assumptions, as raw data and DD methods are typically strictly protected by the owner in practice. In contrast, our threat model adopts a more practical and relaxed assumption, not requiring all users to be benign and permitting the attacker to operate without access to the raw data.

\noindent\textbf{Attack Scenario.} In our threat model, the attacker intercepts the distribution process and injects backdoor information into the benign distilled dataset. The compromised dataset is then redistributed to users, allowing the attacker to manipulate the behavior of downstream models trained on the malicious dataset. 

\noindent\textbf{Attacker’s Goal.}
The primary goal of the attacker is to inject a backdoor into the distilled dataset, ensuring that downstream models trained on it exhibit malicious behavior when triggered, while maintaining high performance on benign inputs.

\noindent\textbf{Attacker’s Capability.}
Our threat model imposes significant constraints on attackers. They do not have access to the raw dataset and can only interact with the distilled dataset, with no prior knowledge of the specific DD method used to generate it.

\noindent\textbf{Challenges.}
\textit{\textbf{i) No Access to Raw Data:}} The attacker has no access to the raw dataset and must infer meaningful information solely from the significantly smaller distilled dataset, often less than one percent of the raw dataset's size.
\textit{\textbf{ii) Bridging the Gap Between Synthetic and Real Images:}} The distilled dataset is highly abstract and lacks the low-level visual details present in the raw data. The attacker must ensure that the injected backdoors are reliably triggered by real-world images in downstream tasks.
\textit{\textbf{iii) Maintaining Dataset Utility:}} The modified distilled dataset must remain effective for training models on legitimate tasks, ensuring the backdoor injection does not degrade overall performance.

\begin{figure*}[!h]
  \centering
  \includegraphics[width=\linewidth]{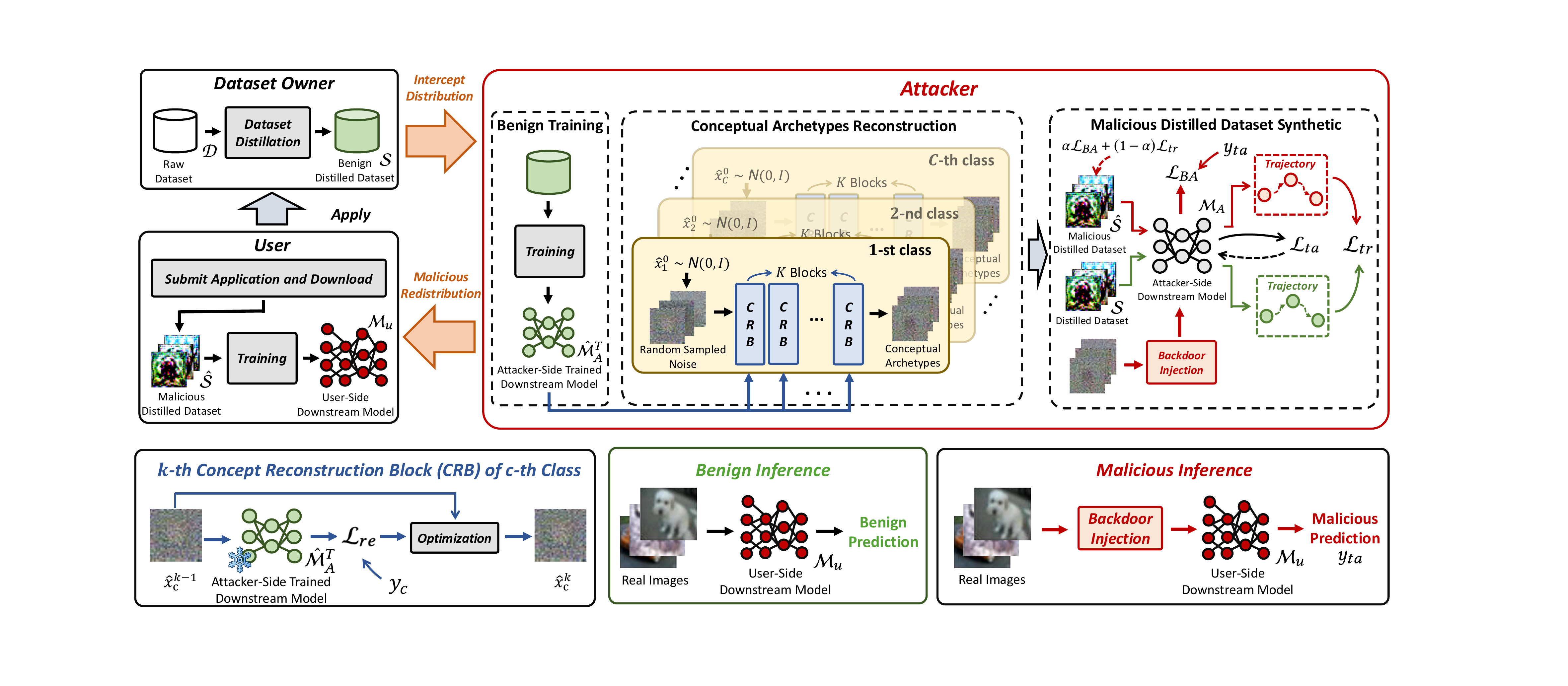}
  \vspace{-15pt}
  \caption{Overview of the proposed method.}
  \vspace{-10pt}
  \label{fig:overview}
\end{figure*}
\section{Proposed Method}
\subsection{Problem Statement}
As mentioned earlier, DD aims to extract knowledge from a large-scale dataset and construct a much smaller synthetic dataset, where models trained on it perform similarly to those trained on the raw dataset. Let $\mathcal{T}$ denote the target dataset and $\mathcal{S}$ the synthetic~(distilled) dataset, where $|\mathcal{T}| \gg |\mathcal{S}|$, indicating that the distilled dataset is much smaller than the original. The loss between the prediction and ground truth is defined as $\ell$. The DD process can then be formulated as~\cite{lei2024tpami}:

\begin{equation}
    \mathbb{E}_{(x, y) \sim \mathcal{D}}\left[ \ell \left( \mathcal{M}_{\mathcal{T}}(x), y \right) \right] \simeq \mathbb{E}_{(x, y) \sim \mathcal{D}} \left[ \ell \left( \mathcal{M}_{\mathcal{S}}(x), y \right) \right],
    \label{eq:ass_1}
\end{equation}
where $\mathcal{M}_{\mathcal{T}}$ and $\mathcal{M}_{\mathcal{S}}$ denote the downstream model $\mathcal{M}$ trained on $\mathcal{T}$ and $\mathcal{S}$, respectively. $\mathcal{D}$ denotes the real data distribution.

In this paper, we aim to update $\mathcal{S}$ to obtain a malicious synthetic dataset $\hat{\mathcal{S}}$, which is injected with backdoor information. The goal is to ensure that malicious behavior is effectively triggered when a model is trained on $\hat{\mathcal{S}}$. The process can be formulated as:
\begin{equation}
\mathbb{E}_{x \sim \mathcal{D}} \left[ \mathcal{M}_{\hat{\mathcal{S}}}(x + T) \right] \approx y_T,
\end{equation}
where $T$ is the trigger and $y_T$ denotes the target label.  
\begin{equation}
\alpha \mathcal{L}_{BA} + (1-\alpha) \mathcal{L}_{tr}. 
\end{equation}

Furthermore, for benign samples, the performance gap between models trained on $\mathcal{S}$ and $\hat{\mathcal{S}}$ should remain minimal to conceal the malicious behavior. The problem can be formulated as:
\begin{equation}
    \mathbb{E}_{(x, y) \sim \mathcal{D}}\left[ \ell \left( \mathcal{M}_{\hat{\mathcal{S}}}(x), y \right) \right] \simeq \mathbb{E}_{(x, y) \sim \mathcal{D}} \left[ \ell \left( \mathcal{M}_{\mathcal{S}}(x), y \right) \right].
    \label{eq:ass_1}
\end{equation}


\subsection{Overview}
The overview of the proposed method is illustrated in Figure~\ref{fig:overview}. As described earlier, our threat model involves three entities: the dataset owner, the attacker, and the benign user. The dataset owner generates a benign distilled dataset $\mathcal{S}$ from the raw dataset $\mathcal{D}$ and distributes it to users upon request. The attacker intercepts the distribution process and converts the benign distilled dataset into a malicious version. 

Specifically, our attack method consists of three main phases. First, the attacker trains a downstream model using the benign distilled dataset $\mathcal{S}$. Next, leveraging the trained model, the attacker reconstructs conceptual archetypes for each class using the proposed Concept Reconstruction Blocks (CRBs). Finally, the attacker injects backdoor information into reconstructed conceptual archetypes and employs a hybrid loss to update the distilled dataset, ensuring that the backdoor is embedded while minimizing performance degradation. Once the malicious distilled dataset is created, it is redistributed to users.

The benign user then trains the local model $\mathcal{M}_u$. Finally, the attacker can target the user-side system by injecting the triggers into real images, activating the malicious behavior in $\mathcal{M}_u$.

\subsection{Proposed Attack Method}

Our attack method consists of three main phases, which work together to effectively inject backdoor information while preserving the knowledge from the raw dataset. We detail each phase in the following sections:

\noindent\textbf{Benign Training.} After intercepting the distribution, the attacker first trains a benign downstream model using the distributed distilled dataset. The attacker-side trained downstream model is defined as $\hat{\mathcal{M}}^T_A$, which is the foundation of the subsequent phases.

\noindent\textbf{Conceptual Archetypes Reconstruction.} Under our strict assumption, the attacker has no access to real images and can only leverage the distilled dataset. However, during the inference phase, the system's input typically consists of real images. This raises a critical question: \textit{\textbf{How can the backdoor be activated when injected into real images without relying on any raw data during backdoor training?}} 

To bridge the gap between distilled and real data, we propose reconstructing conceptual archetypes for each class. Although generating low-level, semantically similar images without access to raw data is infeasible, this limitation is not critical. In deep networks, accurate classification primarily relies on ensuring that the latent feature representations of the conceptual archetypes closely align with those of the real images.

\begin{figure}[!t]
  \centering
  \includegraphics[width=.85\linewidth]{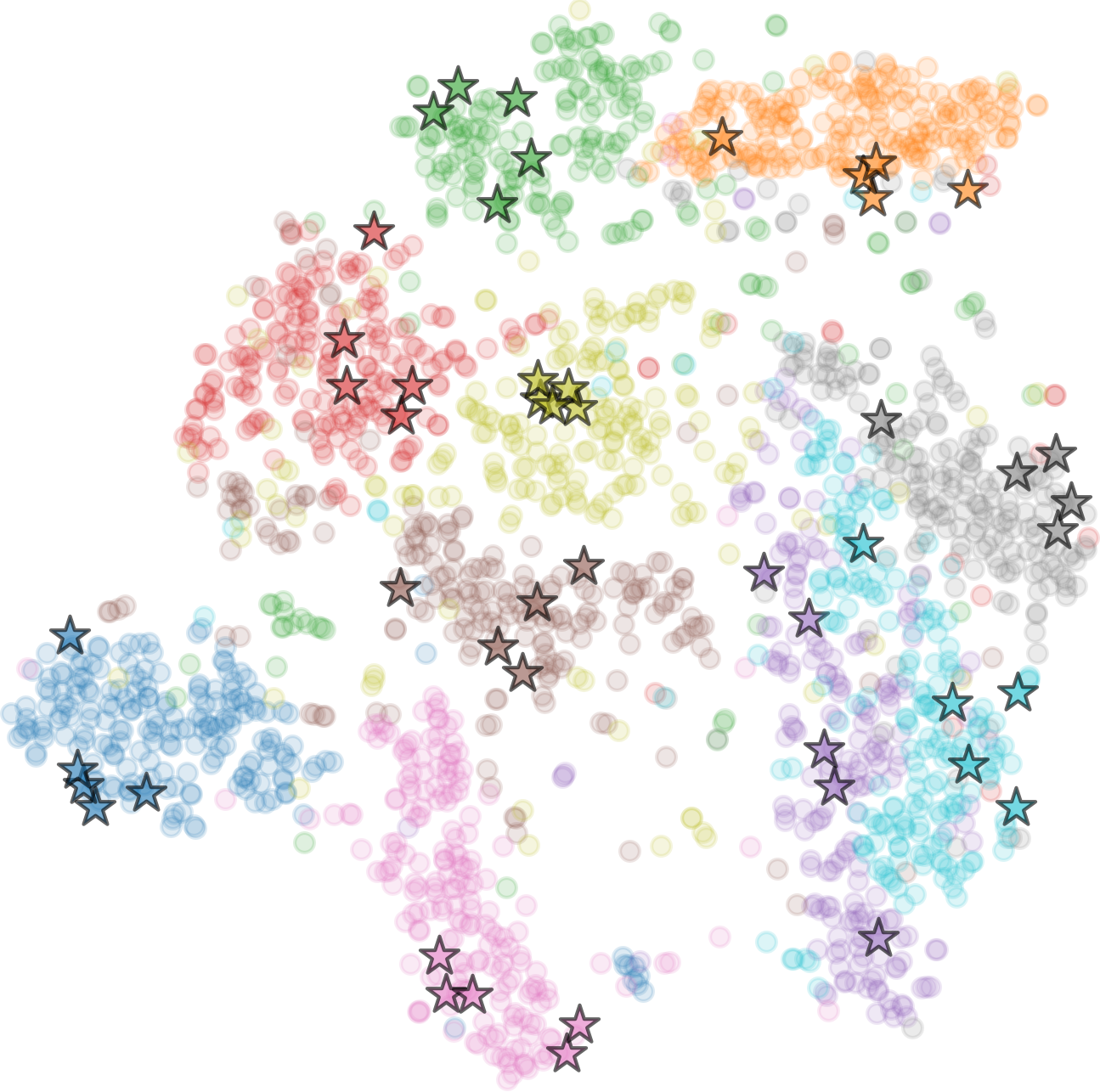}
  \caption{t-SNE visualization of the feature space. ``Stars" and ``Circles" represent the concept archetypes and real images, respectively. The reconstructed archetypes align closely with the deep feature representations of real images, effectively bridging the gap between the distilled data and real images.}
  \vspace{-10pt}
  \label{fig:tsne}
\end{figure}

The reconstruction process aims to generate conceptual archetypes for each class by iteratively refining random noise to align with the high-level feature representations of the target class in $\hat{\mathcal{M}}^T_A$. Specifically, for the $c$-th class, the process consists of $K$ \textbf{Concept Reconstruction Blocks (CRBs)}, each corresponding to an optimization step. The conceptual archetype initialization process for each class $c$ can be formulated as:

\begin{equation}
    \hat{x}_c^{0} \sim \mathcal{N}(0, I),
\end{equation}
where $\mathcal{N}(0, I)$ represents a Gaussian distribution with zero mean and identity covariance matrix. $\hat{x}_c^{0}\in \mathbb{R}^{C\times H\times W}$ denotes the initialized conceptual archetype for the $c$-th class, where $C$, $H$, and $W$ denote the channel, height, and width of the distilled data, respectively.

In the $k$-th CRB block, $\hat{x}_c^{k-1}$ is optimized to align the model's output with the $c$-th class representation. The optimization objective is defined as follows:
\begin{equation}
\mathcal{L}_{\text{re}}(\hat{x}_c^{k-1}, c) = 
- y_c \log \left( \hat{\mathcal{M}}^T_A(\hat{x}_c^{k-1})_c \right),
\end{equation}
where $\mathcal{L}_{\text{re}}$ is the reconstruction loss, $y_c$ represents the one-hot encoded label for class $c$.

The optimization process can be formulated as:
\begin{equation}
\hat{x}_c^{k} = \hat{x}_c^{k-1} - \eta \cdot \nabla_{\hat{x}_c^{k-1}} \mathcal{L}_{\text{re}}(\hat{x}_c^{k-1}, c),
\end{equation}
where $\eta$ is the learning rate, and $\nabla_{\hat{x}_c^{k-1}} \mathcal{L}_{\text{re}}(\hat{x}_c^{k-1}, c)$ represents the gradient of the reconstruction loss with respect to the input $\hat{x}_c^{k-1}$.


After $K$ iterations, the reconstructed image $\hat{x}_c^{K}$ serves as the conceptual archetype for class $c$. This process is repeated $m$ times for each class to generate $m$ archetypes, with $m$ set to 5 in this paper. Figure~\ref{fig:tsne} illustrates a t-SNE visualization~\cite{van2008visualizing} comparing the deep feature representations of the conceptual archetypes with those of real images in MNIST~\cite{lecun1998gradient}. The results show that the reconstructed archetypes closely align with the deep feature representations of real images, effectively bridging the gap between the distilled dataset and real images.

\noindent\textbf{Malicious Distilled Dataset Synthesis.} 
The goal of the attack is to synthesize a malicious distilled dataset such that the backdoor can be effectively activated by real images while maintaining the utility of the dataset for benign tasks. By reconstructing conceptual archetypes to bridge the gap between the distilled and real data, we can leverage them to embed malicious knowledge into the distilled dataset.

Specifically, for each conceptual archetype $\hat{x}$, we obtain the backdoored sample $\hat{x}'$ as follows:
\begin{equation}
\hat{x}'(h, w) =
\begin{cases} 
v, & \text{if } h \geq H - t \, \text{and } w \geq W - t, \\
\hat{x}(h, w), & \text{otherwise},
\end{cases}
\label{eq:inject}
\end{equation}
where  $v$ represents the trigger value and $t$ specifies the trigger size.

Then, a backdoor loss is designed to embed malicious information into the distilled dataset, ensuring that the backdoor behavior is learned by the model trained on the modified data. The backdoor loss is defined as:
\begin{equation}
    \mathcal{L}_{BA}= - y_{ta} \log \left( \mathcal{M}_A(\hat{x}')_{ta} \right),
\end{equation}
where $y_{ta}$ represents the backdoor target label, and $\mathcal{M}_A$ is the attacker-side model, trained from scratch.

To conceal the malicious behavior from detection, it is essential to minimize performance degradation. This requires ensuring that the optimization trajectory of downstream models trained on the malicious distilled dataset closely aligns with those trained on the benign distilled dataset. Specifically, a trajectory consistency loss is introduced to enforce this alignment as follows:
\begin{equation}
\mathcal{L}_{tr} = \frac{1}{|\Theta|} \sum_{\theta \in \Theta} \|\nabla_\theta \mathcal{L}_{ta}(\mathcal{S}) - \nabla_\theta \mathcal{L}_{ta}(\hat{\mathcal{S}})\|^2,
\end{equation}
where $\Theta$ denotes the set of model parameters of $\mathcal{M}_A$, $\mathcal{L}_{ta}$ represents the loss of the downstream task.

By constraining $\mathcal{L}_{tr}$, we can ensure that the malicious dataset maintains a similar optimization trajectory to the benign dataset, thereby concealing malicious behavior while minimizing the impact on the performance of downstream tasks. Finally, we combine both losses to form the overall objective for synthesizing the malicious distilled dataset. The hybrid loss function is defined as follows:
\begin{equation}
    \mathcal{L}_{hybrid} = \alpha \mathcal{L}_{BA} + (1 - \alpha) \mathcal{L}_{tr},
    \label{eq:loss}
\end{equation}
where $\alpha$ is the balancing parameter that controls the trade-off between embedding malicious information and maintaining trajectory consistency.

Then, $\hat{\mathcal{S}}$ is iteratively updated to minimize $\mathcal{L}_{hybrid}$ as:
\begin{equation}
    \hat{\mathcal{S}} \leftarrow \hat{\mathcal{S}} - \eta \cdot \nabla \mathcal{L}_\text{hybrid}.
\end{equation}


These steps are repeated for $N$ iterations within a single epoch. To ensure that model $\mathcal{M}_A$ follows the next benign optimization trajectory, it is updated on $\mathcal{S}$ after each epoch. This entire process is repeated for $E$ epochs.

\noindent\textbf{Implementation.} Once the attacker synthesizes the malicious distilled dataset \(\hat{\mathcal{S}}\), it is redistributed to the users. Users then train their downstream models \(\mathcal{M}_u\) on \(\hat{\mathcal{S}}\) using their own training strategies. 

During the inference phase, the malicious behavior is activated when the trigger is injected into real images following Eq.~\eqref{eq:inject} to produce malicious outputs aligned with the attacker's target, while maintaining normal performance on benign inputs.

Notably, our attack method remains effective even when $\mathcal{M}_u$ and $\mathcal{M}_A$ have different architectures. Furthermore, it does not require fine-tuning any DD process on the dataset owner’s side, nor does it require access to raw data. Therefore, our method is versatile and practical across various scenarios.

\begin{table*}[!t]
\caption{Experimental results of different strategies and different raw datasets with DC (Avg $\pm$ STD, \%).}
\vspace{-10pt}
\label{tab:stra}
\centering
\resizebox{\textwidth}{!}{
\begin{tabular}{c|c|p{1.2cm}|>{\centering\arraybackslash}p{1.7cm}>{\centering\arraybackslash}p{1.7cm}>{\centering\arraybackslash}p{1.7cm}>{\centering\arraybackslash}p{1.7cm}>{\centering\arraybackslash}p{1.7cm}>{\centering\arraybackslash}p{1.7cm}>{\centering\arraybackslash}p{1.7cm}>{\centering\arraybackslash}p{1.7cm}>{\centering\arraybackslash}p{1.7cm}>{\centering\arraybackslash}p{1.7cm}}
\hline
\multicolumn{1}{l|}{} & IPC   &   \diagbox[width=1.6cm]{{\small Metric}}{{\small Epoch}}       & 10 & 20 & 30 & 40 & 50 & 60 & 70 & 80 & 90 & 100 \\
\hline 
\multirow{6}{*}{\rotatebox{90}{CIFAR10}} & \multirow{3}{*}{1} & Baseline & 26.88$\pm$0.405   &  27.25$\pm$0.500  & 27.56$\pm$0.481   &   27.85$\pm$0.428 & 27.78$\pm$0.592   & 27.85$\pm$0.526   &  27.83$\pm$0.631  &   27.45$\pm$0.607 &  27.52$\pm$0.753  &  27.77$\pm$0.383   \\
                        & & BA            & 25.07$\pm$0.475
    & 25.47$\pm$0.610
    & 25.42$\pm$0.574
    & 25.50$\pm$0.579
    & 25.43$\pm$0.415
    & 25.82$\pm$0.649
    & 25.31$\pm$0.611
    & 25.71$\pm$0.508
    & 25.32$\pm$0.521
    & 25.79$\pm$0.448 \\ 
                      &  & \cellcolor{lightred}ASR     & \cellcolor{lightred}99.82$\pm$0.150
    & \cellcolor{lightred}99.99$\pm$0.009
    & \cellcolor{lightred}100.00$\pm$0.000
    & \cellcolor{lightred}99.99$\pm$0.013
    & \cellcolor{lightred}100.00$\pm$0.006
    & \cellcolor{lightred}100.00$\pm$0.000
    & \cellcolor{lightred}100.00$\pm$0.000
    & \cellcolor{lightred}100.00$\pm$0.003
    & \cellcolor{lightred}100.00$\pm$0.006
    &\cellcolor{lightred} 100.00$\pm$0.003 \\ \cline{2-13}
 & \multirow{3}{*}{10} & Baseline &    26.57$\pm$0.996
 &  31.85$\pm$0.836  &  35.62$\pm$0.466  &  38.40$\pm$0.447  & 39.93$\pm$0.760   &  40.86$\pm$0.484  &  41.49$\pm$0.620  &  41.99$\pm$0.441  & 42.77$\pm$0.745   & 42.63$\pm$0.426    \\
    & & BA          & 25.82$\pm$0.749
    & 30.22$\pm$0.611
    & 32.59$\pm$0.540
    & 34.07$\pm$0.391
    & 34.87$\pm$0.386
    & 35.24$\pm$0.417
    & 35.32$\pm$0.585
    & 35.98$\pm$0.491
    & 35.60$\pm$0.376
    & 35.92$\pm$0.442
   \\ 
    &  & \cellcolor{lightred} ASR     & \cellcolor{lightred}99.99$\pm$0.008
    & \cellcolor{lightred}100.00$\pm$0.000
    & \cellcolor{lightred}100.00$\pm$0.000
    & \cellcolor{lightred}100.00$\pm$0.000
    & \cellcolor{lightred}100.00$\pm$0.000
    & \cellcolor{lightred}100.00$\pm$0.000
    & \cellcolor{lightred}99.99$\pm$0.013
    & \cellcolor{lightred}100.00$\pm$0.009
    & \cellcolor{lightred}100.00$\pm$0.012
    & \cellcolor{lightred}100.00$\pm$0.003 \\
[0.5ex] \hline \hline

\multirow{3}{*}{\rotatebox{90}{CIFAR100}} & \multirow{3}{*}{1} & Baseline &  3.36$\pm$0.270  & 6.13$\pm$0.264   &  7.99$\pm$0.548  &  9.05$\pm$0.471  &  9.46$\pm$0.294  &   10.22$\pm$0.380 &  10.52$\pm$0.237  & 10.69$\pm$0.335   &  11.11$\pm$0.318  & 10.77$\pm$0.294    \\ [0.5ex]
                        & & BA         & 3.05$\pm$0.338
    & 5.95$\pm$0.319
    & 7.24$\pm$0.317
    & 8.36$\pm$0.418
    & 8.59$\pm$0.138
    & 9.19$\pm$0.190
    & 9.39$\pm$0.220
    & 9.75$\pm$0.204
    & 9.81$\pm$0.268
    & 9.97$\pm$0.374   \\ [0.5ex]
                      &  &\cellcolor{lightred} ASR     & \cellcolor{lightred}22.56$\pm$27.631
    & \cellcolor{lightred}87.38$\pm$10.855
    & \cellcolor{lightred}96.93$\pm$6.836
    & \cellcolor{lightred}98.89$\pm$0.849
    & \cellcolor{lightred}99.26$\pm$1.155
    & \cellcolor{lightred}99.80$\pm$0.285
    & \cellcolor{lightred}99.49$\pm$0.730
    & \cellcolor{lightred}99.85$\pm$0.160
    & \cellcolor{lightred}99.57$\pm$0.600
    & \cellcolor{lightred}99.83$\pm$0.153    \\[0.5ex]

   \cline{2-13}
  
\hline \hline
\multirow{9}{*}{\rotatebox{90}{MNIST}} & \multirow{3}{*}{1} & Baseline & 70.69$\pm$2.768   &  79.73$\pm$2.424  & 82.19$\pm$0.874 &  84.91$\pm$1.659  & 85.92$\pm$0.944   & 86.38$\pm$1.227   & 86.75$\pm$0.730   & 87.27$\pm$0.737   &  87.89$\pm$0.553  &  88.81$\pm$0.867   \\
                        & & BA        & 65.69$\pm$2.740
    & 70.42$\pm$2.298
    & 73.91$\pm$2.185
    & 75.04$\pm$2.094
    & 76.65$\pm$1.384
    & 77.00$\pm$1.612
    & 77.55$\pm$1.518
    & 78.38$\pm$1.458
    & 78.75$\pm$1.235
    & 79.45$\pm$1.073   \\
                      &  &\cellcolor{lightred} ASR     & \cellcolor{lightred}100.00$\pm$0.000
    & \cellcolor{lightred}100.00$\pm$0.000
    & \cellcolor{lightred}100.00$\pm$0.000
    & \cellcolor{lightred}100.00$\pm$0.000
    & \cellcolor{lightred}100.00$\pm$0.000
    & \cellcolor{lightred}100.00$\pm$0.000
    & \cellcolor{lightred}100.00$\pm$0.000
    & \cellcolor{lightred}100.00$\pm$0.000
    & \cellcolor{lightred}100.00$\pm$0.000
    & \cellcolor{lightred}100.00$\pm$0.000     \\ \cline{2-13}
& \multirow{3}{*}{10} & Baseline &  69.75$\pm$5.173  &  80.26$\pm$2.411  &  83.67$\pm$0.885  &  86.17$\pm$0.703  &  89.32$\pm$0.510  &  91.49$\pm$0.471  &  93.23$\pm$0.246  & 94.63$\pm$0.287   &  95.17$\pm$0.170  &  95.72$\pm$0.232   \\
    & & BA         & 63.76$\pm$3.923
    & 75.31$\pm$1.948
    & 79.98$\pm$1.134
    & 82.05$\pm$1.565
    & 85.09$\pm$1.058
    & 87.51$\pm$0.817
    & 89.35$\pm$0.661
    & 90.07$\pm$0.458
    & 90.39$\pm$1.048
    & 90.41$\pm$0.478  \\
   &  &\cellcolor{lightred} ASR     & \cellcolor{lightred}100.00$\pm$0.000
    & \cellcolor{lightred}100.00$\pm$0.000
    & \cellcolor{lightred}100.00$\pm$0.000
    & \cellcolor{lightred}100.00$\pm$0.000
    & \cellcolor{lightred}100.00$\pm$0.000
    & \cellcolor{lightred}100.00$\pm$0.000
    & \cellcolor{lightred}100.00$\pm$0.000
    & \cellcolor{lightred}100.00$\pm$0.000
    & \cellcolor{lightred}100.00$\pm$0.000
    & \cellcolor{lightred}100.00$\pm$0.000     \\ \cline{2-13}
& \multirow{3}{*}{50} & Baseline &  78.09$\pm$1.537  &  85.41$\pm$0.717  &  90.01$\pm$0.653  &93.57$\pm$0.293& 95.09$\pm$0.154   & 95.98$\pm$0.189   & 96.69$\pm$0.110   & 97.13$\pm$0.140   &  97.44$\pm$0.111  &  97.74$\pm$0.072   \\
    & & BA         & 64.23$\pm$4.176
    & 68.67$\pm$1.887
    & 74.48$\pm$1.117
    & 79.52$\pm$1.135
    & 85.17$\pm$1.045
    & 87.25$\pm$0.670
    & 88.49$\pm$0.663
    & 88.94$\pm$0.540
    & 89.32$\pm$0.466
    & 89.31$\pm$0.625
    \\
    &  & \cellcolor{lightred} ASR      & \cellcolor{lightred}100.00$\pm$0.000
    & \cellcolor{lightred}100.00$\pm$0.000
    & \cellcolor{lightred}100.00$\pm$0.000
    & \cellcolor{lightred}100.00$\pm$0.000
    & \cellcolor{lightred}100.00$\pm$0.000
    & \cellcolor{lightred}100.00$\pm$0.000
    & \cellcolor{lightred}100.00$\pm$0.000
    & \cellcolor{lightred}100.00$\pm$0.000
    & \cellcolor{lightred}100.00$\pm$0.000
    & \cellcolor{lightred}100.00$\pm$0.000  
     \\
\hline \hline
\multirow{9}{*}{\rotatebox{90}{FashionMNIST}} & \multirow{3}{*}{1} & Baseline   & 67.51$\pm$0.399
  & 69.29$\pm$0.741
  & 69.45$\pm$0.908
  & 69.81$\pm$0.843
  & 69.88$\pm$0.715
  & 70.10$\pm$0.787
  & 69.94$\pm$0.553
  & 69.86$\pm$0.553
  & 69.84$\pm$0.525
  & 69.98$\pm$0.680    \\
                        & & BA           & 61.75$\pm$1.044
    & 63.29$\pm$0.741
    & 63.72$\pm$0.803
    & 63.66$\pm$0.870
    & 63.66$\pm$0.382
    & 63.70$\pm$0.661
    & 63.37$\pm$0.755
    & 63.93$\pm$0.933
    & 63.94$\pm$0.786
    & 63.93$\pm$0.553  \\   &  & \cellcolor{lightred} ASR      & \cellcolor{lightred}100.00$\pm$0.000
    & \cellcolor{lightred}100.00$\pm$0.000
    & \cellcolor{lightred}100.00$\pm$0.000
    & \cellcolor{lightred}100.00$\pm$0.000
    & \cellcolor{lightred}100.00$\pm$0.000
    & \cellcolor{lightred}100.00$\pm$0.000
    & \cellcolor{lightred}100.00$\pm$0.000
    & \cellcolor{lightred}100.00$\pm$0.000
    & \cellcolor{lightred}100.00$\pm$0.000
    & \cellcolor{lightred}100.00$\pm$0.000    \\ \cline{2-13}
& \multirow{3}{*}{10} & Baseline & 60.92$\pm$1.977
  & 66.27$\pm$1.095
  & 69.41$\pm$0.766
  & 72.39$\pm$0.445
  & 74.29$\pm$0.468
  & 75.51$\pm$0.296
  & 76.41$\pm$0.278
  & 77.61$\pm$0.393
  & 78.15$\pm$0.176
  & 78.81$\pm$0.209     \\
    & & BA           & 59.84$\pm$4.410
    & 65.94$\pm$1.568
    & 69.02$\pm$1.264
    & 69.83$\pm$0.799
    & 71.02$\pm$0.561
    & 71.12$\pm$0.482
    & 70.89$\pm$0.744
    & 71.36$\pm$0.738
    & 71.17$\pm$0.533
    & 71.42$\pm$0.633  \\
    &   & \cellcolor{lightred} ASR      & \cellcolor{lightred}100.00$\pm$0.000
    & \cellcolor{lightred}100.00$\pm$0.000
    & \cellcolor{lightred}100.00$\pm$0.000
    & \cellcolor{lightred}100.00$\pm$0.000
    & \cellcolor{lightred}100.00$\pm$0.000
    & \cellcolor{lightred}100.00$\pm$0.000
    & \cellcolor{lightred}100.00$\pm$0.000
    & \cellcolor{lightred}100.00$\pm$0.000
    & \cellcolor{lightred}100.00$\pm$0.000
    & \cellcolor{lightred}100.00$\pm$0.000  \\ \cline{2-13}
& \multirow{3}{*}{50} & Baseline   & 65.95$\pm$0.638
  & 69.32$\pm$0.390
  & 71.54$\pm$0.273
  & 72.94$\pm$0.279
  & 74.73$\pm$0.289
  & 76.17$\pm$0.273
  & 77.16$\pm$0.325
  & 77.85$\pm$0.213
  & 78.71$\pm$0.253
  & 79.40$\pm$0.148 \\
    & & BA         & 63.04$\pm$0.868
    & 67.18$\pm$0.499
    & 69.22$\pm$0.260
    & 69.95$\pm$0.454
    & 69.96$\pm$0.268
    & 70.01$\pm$0.350
    & 69.32$\pm$0.347
    & 69.58$\pm$0.447
    & 69.37$\pm$0.353
    & 69.49$\pm$0.704 \\
    &  & \cellcolor{lightred} ASR      & \cellcolor{lightred}100.00$\pm$0.000
    & \cellcolor{lightred}100.00$\pm$0.000
    & \cellcolor{lightred}100.00$\pm$0.000
    & \cellcolor{lightred}100.00$\pm$0.000
    & \cellcolor{lightred}100.00$\pm$0.000
    & \cellcolor{lightred}99.94$\pm$0.089
    & \cellcolor{lightred}99.93$\pm$0.059
    & \cellcolor{lightred}99.64$\pm$0.434
    & \cellcolor{lightred}99.58$\pm$0.434
    & \cellcolor{lightred}99.51$\pm$0.461 \\\hline \hline
\multirow{3}{*}{\rotatebox{90}{SVHN}} & \multirow{3}{*}{1} & Baseline   & 29.10$\pm$1.480
  & 31.58$\pm$1.154
  & 31.37$\pm$1.511
  & 30.63$\pm$1.027
  & 29.91$\pm$0.851
  & 29.90$\pm$0.977
  & 30.51$\pm$1.481
  & 29.47$\pm$0.898
  & 30.20$\pm$1.023
  & 30.64$\pm$1.842    \\
                        & & BA          & 29.30$\pm$1.276
    & 30.59$\pm$0.979
    & 29.70$\pm$1.376
    & 29.05$\pm$1.842
    & 28.69$\pm$1.322
    & 29.57$\pm$1.004
    & 28.88$\pm$1.080
    & 28.99$\pm$1.718
    & 29.56$\pm$0.957
    & 29.47$\pm$0.926  \\
                      &  &\cellcolor{lightred} ASR      & \cellcolor{lightred}100.00$\pm$0.000
    & \cellcolor{lightred}100.00$\pm$0.000
    & \cellcolor{lightred}100.00$\pm$0.000
    & \cellcolor{lightred}100.00$\pm$0.000
    & \cellcolor{lightred}100.00$\pm$0.000
    & \cellcolor{lightred}100.00$\pm$0.000
    & \cellcolor{lightred}100.00$\pm$0.000
    & \cellcolor{lightred}100.00$\pm$0.000
    & \cellcolor{lightred}100.00$\pm$0.000
    & \cellcolor{lightred}100.00$\pm$0.000    
    \\ \hline

\end{tabular}
}
\vspace{-10pt}
\end{table*}

\section{Experiments}
\subsection{Experimental Setting}
\noindent \textbf{Experiment Environment.}
Our proposed method is implemented using the PyTorch framework and optimized with Stochastic Gradient Descent~(SGD)~\cite{kingma2014adam} with a learning rate of 0.01. The number of epochs for synthesizing the malicious dataset is set to 10. The experiments are conducted on a system with an AMD Ryzen 7 5800X CPU @3.80 GHz, 32 GB of RAM, and an NVIDIA GTX 3090.

\noindent \textbf{Experiment Setting.} In our experiments, we use ConvNet~\cite{krizhevsky2012imagenet} as the default attacker-side downstream model. Additionally,  AlexNet~\cite{krizhevsky2017imagenet}, VGG11~\cite{simonyan2014very}, VGG16~\cite{simonyan2014very}, ResNet18~\cite{he2016deep}, and ResNet34~\cite{he2016deep} are used as the user-side downstream networks.

\noindent \textbf{Datasets.} To assess the generalization ability of our method across different raw datasets, we evaluate it on CIFAR-10~\cite{krizhevsky2009learning}, CIFAR-100~\cite{krizhevsky2009learning}, MNIST~\cite{lecun1998gradient}, FashionMNIST~\cite{xiao2017fashion}, and SVHN~\cite{netzer2011reading}.

\begin{figure*}[!t]
    \centering
    \begin{subfigure}[b]{0.335\textwidth}
        \centering
        \includegraphics[width=\textwidth]{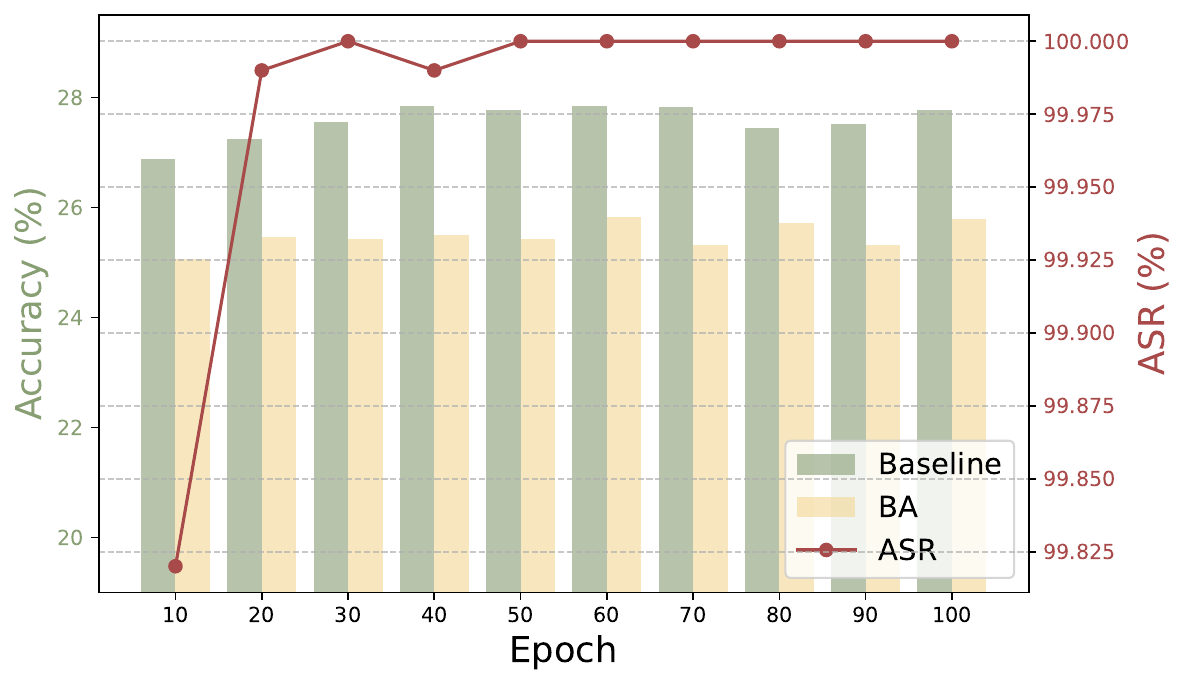}
        \vspace{-15pt}
        \caption{ConvNet}
    \end{subfigure}
    \hfill
    \begin{subfigure}[b]{0.32\textwidth}
        \centering
        \includegraphics[width=\textwidth]{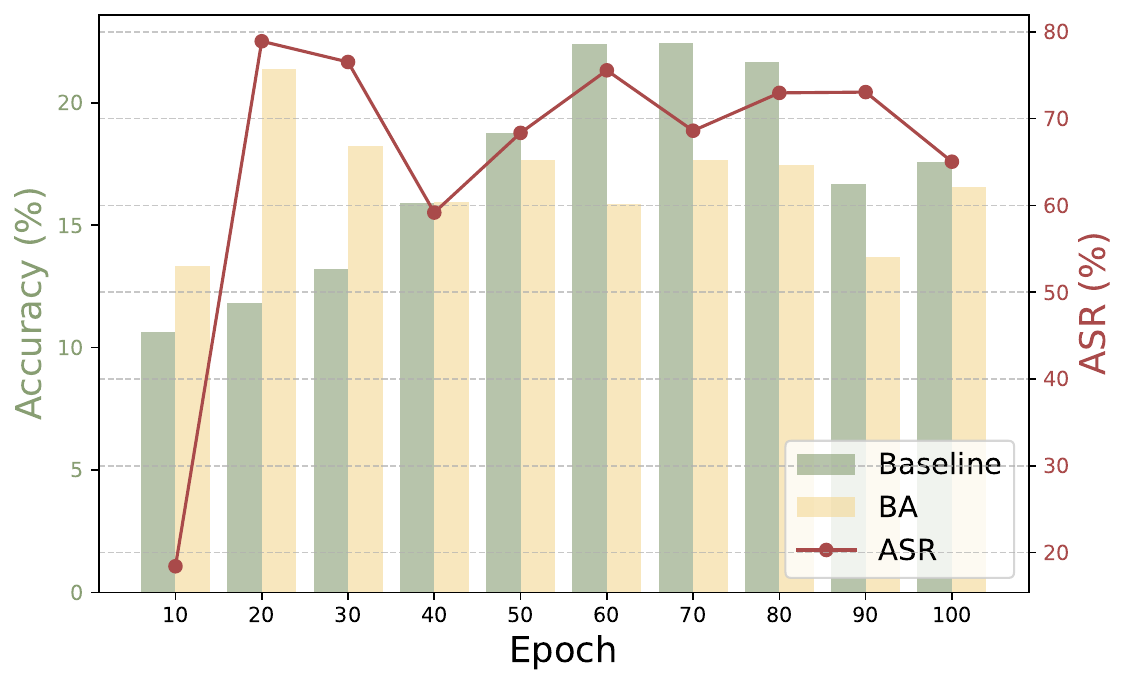}
                \vspace{-15pt}
        \caption{AlexNet}
    \end{subfigure}
    \hfill
    \begin{subfigure}[b]{0.33\textwidth}
        \centering
        \includegraphics[width=\textwidth]{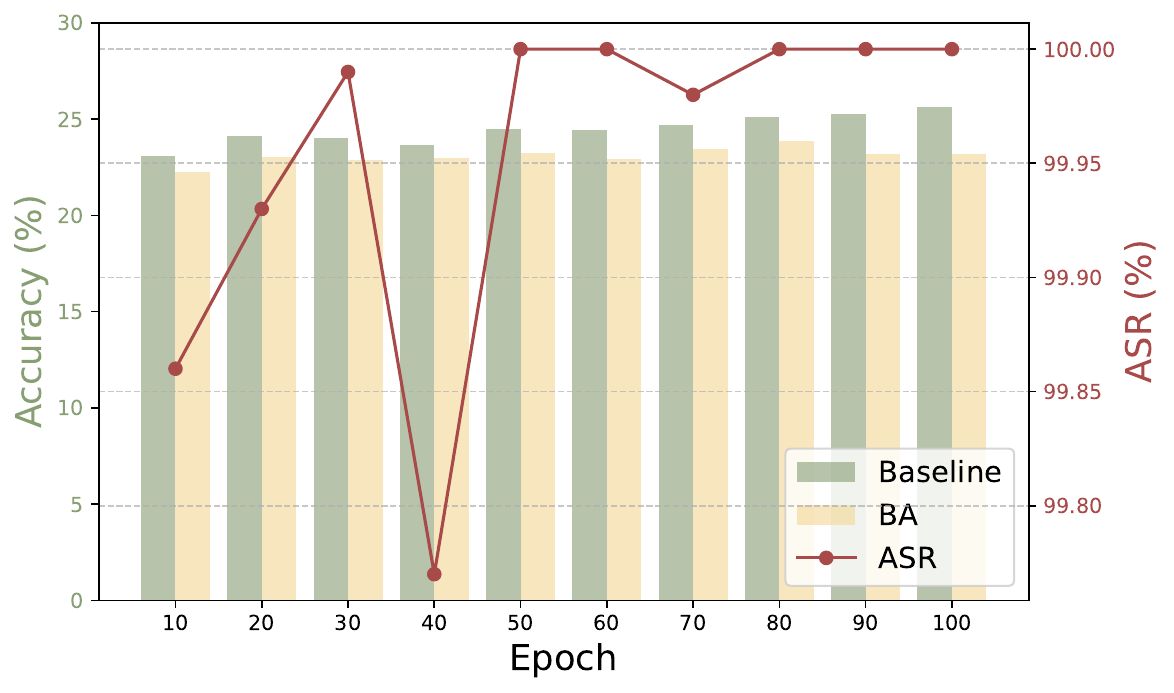}
                \vspace{-15pt}
        \caption{VGG11}
    \end{subfigure}

    \begin{subfigure}[b]{0.33\textwidth}
        \centering
        \includegraphics[width=\textwidth]{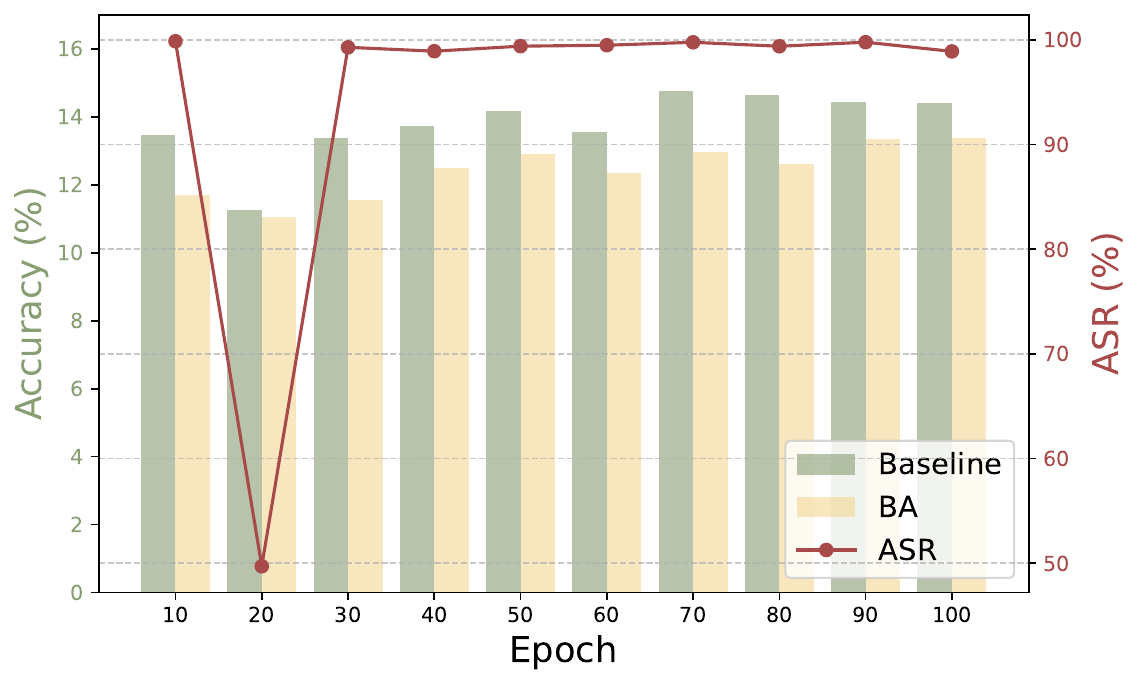}
                \vspace{-15pt}
        \caption{VGG16}
    \end{subfigure}
    \hfill
    \begin{subfigure}[b]{0.33\textwidth}
        \centering
        \includegraphics[width=\textwidth]{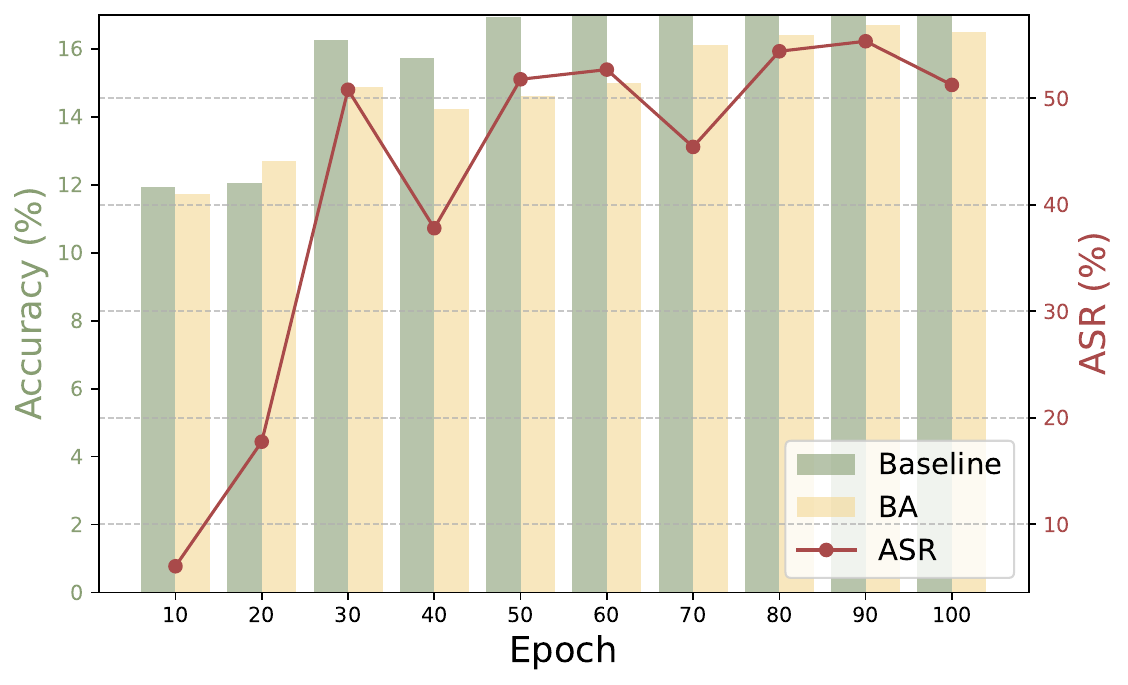}
                \vspace{-15pt}
        \caption{ResNet18}
    \end{subfigure}
    \hfill
    \begin{subfigure}[b]{0.33\textwidth}
        \centering
        \includegraphics[width=\textwidth]{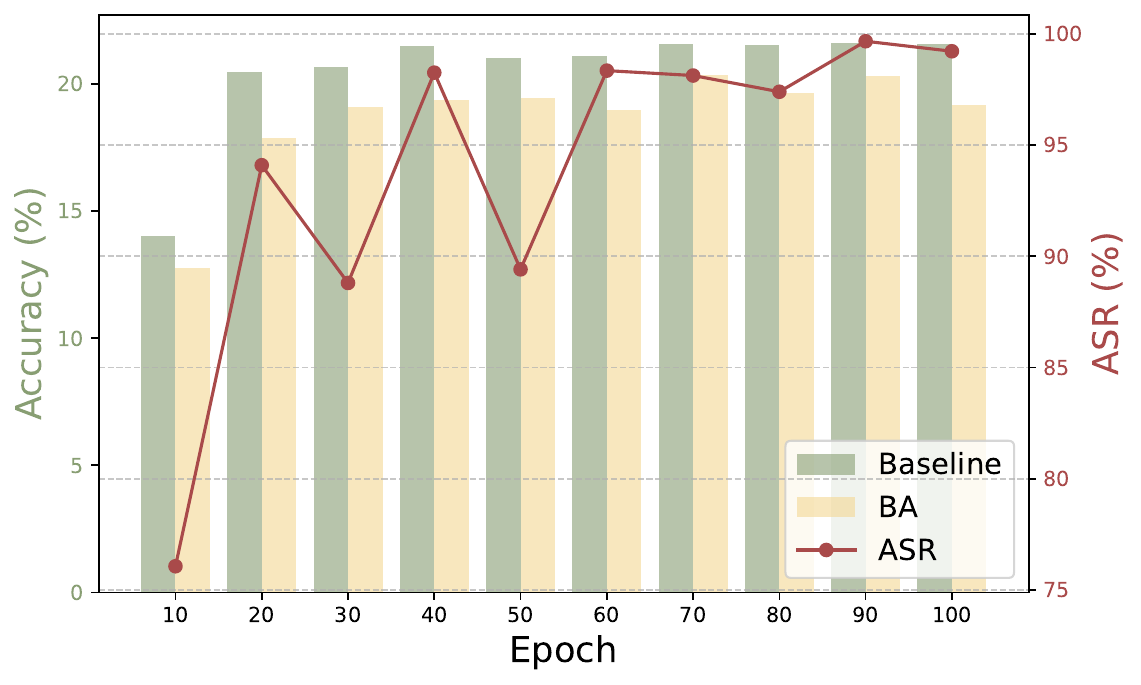}
                \vspace{-15pt}
        \caption{ResNet34}
    \end{subfigure}
    \vspace{-20pt}
    \caption{The performances of different user-side models under different training strategies. Our attack consistently poses a significant threat across different user-side models and training strategies.}
    \vspace{-10pt}
    \label{fig:performance}
\end{figure*}

\noindent \textbf{DD Methods.} To evaluate the generalizability of our method, we test it across several different representative DD methods, including DC~\cite{zhao2021dataset}, DM~\cite{zhao2023DM}, DSA~\cite{zhao2021datasetdsa}, and MTT~\cite{cazenavette2022dataset}.

\noindent \textbf{Metrics.} Similar to other backdoor attack methods~\cite{gu2019badnets}, we adopt benign accuracy (BA) to measure performance on benign samples. Besides, we use attack success rate (ASR) to the effectiveness of our attack. 

Neural networks exhibit inherent randomness due to variations introduced by different random seeds. To ensure the reliability of our results, we conduct experiments with 10 different seeds and report both the average (Avg) and standard deviation (STD) of the performance metrics.

\begin{table*}[!t]
\caption{Experimental results based on different DD methods and different user-side models (Avg $\pm$ STD, \%).}
\vspace{-5pt}
\label{tab:tab2}
\centering
\resizebox{\textwidth}{!}{
\begin{tabular}{c|c|c|p{1cm}|cc|cc|cc|cc|cc|cc}
\hline
 \multirow{2}{*}{Method} & \multirow{2}{*}{Dataset} &  \multirow{2}{*}{IPC}   &    &   \multicolumn{2}{c}{ConvNet}  &   \multicolumn{2}{|c}{AlexNet} & \multicolumn{2}{|c}{VGG11}  & \multicolumn{2}{|c}{VGG16} & \multicolumn{2}{|c}{ResNet18} & \multicolumn{2}{|c}{ResNet34}  \\
& & & & \multicolumn{1}{c}{50} & 100 & \multicolumn{1}{c}{50} & 100 & \multicolumn{1}{c}{50} & 100 & \multicolumn{1}{c}{50} & 100 & \multicolumn{1}{c}{50} & 100 & \multicolumn{1}{c}{50} & 100 
 \\
\hline 
\multirow{12}{*}{DC} & \multirow{6}{*}{\rotatebox{90}{CIFAR10}} & \multirow{3}{*}{1} & Baseline & 27.78$\pm$0.592  & 27.77$\pm$0.383 & 18.77$\pm$0.022 & 17.56$\pm$0.275 & 24.47$\pm$0.921 & 25.66$\pm$0.952 & 14.16$\pm$1.069 & 14.42$\pm$0.958 & 16.96$\pm$1.073 & 17.93$\pm$1.021 & 21.01$\pm$0.886 & 21.58$\pm$0.851\\
                      &  & & BA            & 25.43$\pm$0.415 & 25.79$\pm$0.448 & 17.68$\pm$2.401 & 16.57$\pm$3.968 & 23.24$\pm$0.747 & 23.17$\pm$0.921 & 12.90$\pm$0.866 & 13.37$\pm$1.015 & 14.62$\pm$1.645 & 16.49$\pm$0.970 & 19.45$\pm$1.062 & 19.17$\pm$1.404\\ 
                    &  &  & \cellcolor{lightred}ASR     & \cellcolor{lightred}100.00$\pm$0.006 & \cellcolor{lightred}100.00$\pm$0.003 & \cellcolor{lightred} 68.35$\pm$30.250 & \cellcolor{lightred} 65.03$\pm$33.621& \cellcolor{lightred} 100.00$\pm$0.010& \cellcolor{lightred} 100.00$\pm$0.004& \cellcolor{lightred} 99.40$\pm$1.310& \cellcolor{lightred} 98.90$\pm$2.106& \cellcolor{lightred} 51.77$\pm$39.082& \cellcolor{lightred} 51.24$\pm$26.014& \cellcolor{lightred} 89.41$\pm$18.270& \cellcolor{lightred} 99.21$\pm$1.374
                    \\ \cline{3-16}
& & \multirow{3}{*}{10} & Baseline & 39.93$\pm$0.760 &  42.63$7\pm$0.426  & 12.79$\pm$1.573 & 21.48$\pm$0.729 & 34.67$\pm$0.618 & 35.27$\pm$0.497 & 23.71$\pm$1.556 & 26.16$\pm$1.630 & 17.20$\pm$1.152 & 18.52$\pm$1.199 & 22.08$\pm$1.464 & 22.88$\pm$1.339\\
&    & & BA          &  34.87$7\pm$0.386 &  35.92$7\pm$0.442 & 22.26$\pm$4.296 & 16.38$\pm$4.978 & 28.48$\pm$0.571 & 29.07$\pm$0.903 & 20.76$\pm$0.722 & 22.15$\pm$0.876 & 14.02$\pm$0.665 & 14.25$\pm$0.746  & 17.73$\pm$0.830 & 18.47$\pm$0.951
   \\ 
&    &  & \cellcolor{lightred} ASR     & \cellcolor{lightred}100.00$\pm$0.000  & \cellcolor{lightred}100.00$\pm$0.003 & \cellcolor{lightred}96.84$\pm$6.607 &\cellcolor{lightred} 41.41$\pm$41.240 & \cellcolor{lightred}100.00$\pm$0.000 & \cellcolor{lightred}100.00$\pm$0.000 & \cellcolor{lightred}100.00$\pm$0.000 & \cellcolor{lightred}100.00$\pm$0.000 & \cellcolor{lightred}75.51$\pm$17.345 & \cellcolor{lightred}88.00$\pm$14.494 & \cellcolor{lightred}99.95$\pm$0.131 & \cellcolor{lightred}99.97$\pm$0.096 \\  \cline{2-16} 
 & \multirow{3}{*}{\rotatebox{90}{CIFAR100}} & \multirow{3}{*}{1} & Baseline & 9.46$\pm$0.294 & 10.77$\pm$0.294 & 1.21$\pm$0.197 & 1.68$\pm$0.520 & 8.41$\pm$0.356& 9.01$\pm$0.208 & 3.15$\pm$0.436 & 4.61$\pm$0.599 & 1.53$\pm$0.140 & 1.67$\pm$0.184 & 2.26$\pm$0.214 & 3.16$\pm$0.461
 \\ [0.55ex]
 &    & & BA  & 8.59$\pm$0.138 & 9.97$\pm$0.374 & 2.11$\pm$0.319 & 1.33$\pm$0.401 & 7.06$\pm$0.334 & 7.85$\pm$0.295 & 2.85$\pm$0.742 & 4.09$\pm$0.389 & 1.34$\pm$0.146 & 1.39$\pm$0.166 & 1.71$\pm$0.131 & 2.30$\pm$0.229    \\ [0.55ex]
&    &  & \cellcolor{lightred} ASR   & \cellcolor{lightred}99.26$\pm$1.155 & \cellcolor{lightred}99.83$\pm$0.153 & \cellcolor{lightred}33.27$\pm$27.747  & \cellcolor{lightred}0.00$\pm$0.000 & \cellcolor{lightred}97.67$\pm$3.981 & \cellcolor{lightred}99.17$\pm$2.364 & \cellcolor{lightred}91.98$\pm$23.926 & \cellcolor{lightred}87.85$\pm$24.757 & \cellcolor{lightred}0.40$\pm$0.544 & \cellcolor{lightred}0.03$\pm$0.066 & \cellcolor{lightred}22.94$\pm$23.976 & \cellcolor{lightred}15.05$\pm$22.299 \\ [0.55ex] \cline{2-16}
 & \multirow{6}{*}{\rotatebox{90}{FashionMNIST}} & \multirow{3}{*}{1} &  Baseline & 69.88$\pm$0.715 & 69.98$\pm$0.680 & 52.15$\pm$4.806 & 30.74$\pm$14.513 & 59.75$\pm$4.468 & 62.72$\pm$2.498 & 23.80$\pm$5.390 & 31.16$\pm$5.491 & 57.98$\pm$1.642 & 57.29$\pm$1.721 & 61.80$\pm$1.846 & 61.92$\pm$1.641 \\
 &    & & BA   & 63.66$\pm$0.382 & 63.93$\pm$0.553 & 29.17$\pm$16.787 & 20.30$\pm$14.193 & 55.05$\pm$2.035 & 55.46$\pm$3.141 & 21.60$\pm$4.460 & 27.60$\pm$3.811 & 50.18$\pm$1.812  & 51.59$\pm$1.355 & 54.87$\pm$2.604 & 53.47$\pm$3.275   \\
&    &  & \cellcolor{lightred} ASR   & \cellcolor{lightred}100.00$\pm$0.000 & \cellcolor{lightred}100.00$\pm$0.000 & \cellcolor{lightred}74.46$\pm$40.198 & \cellcolor{lightred}68.27$\pm$41.138 & \cellcolor{lightred}100.00$\pm$0.000 & \cellcolor{lightred}100.00$\pm$0.000 & \cellcolor{lightred}100.00$\pm$0.000 & \cellcolor{lightred}100.00$\pm$0.000 & \cellcolor{lightred}99.66$\pm$0.646 & \cellcolor{lightred}98.43$\pm$4.697 & \cellcolor{lightred}100.00$\pm$0.000 & \cellcolor{lightred}100.00$\pm$0.000 \\
& &    \multirow{3}{*}{10} & Baseline &  74.29$\pm$0.468 & 78.81$\pm$0.209 & 25.84$\pm$3.626 & 53.17$\pm$11.620 & 77.46$\pm$0.596 & 78.00$\pm$0.495 & 56.81$\pm$1.701 & 63.27$\pm$4.275
& 56.49$\pm$1.497 & 57.77$\pm$1.642 & 59.97$\pm$2.190 & 62.40$\pm$1.877\\
 &    & & BA & 71.02$\pm$0.561 & 71.42$\pm$0.633 & 19.12$\pm$9.972 & 34.29$\pm$15.116& 70.53$\pm$0.991& 70.55$\pm$0.985  & 46.80$\pm$4.070  & 55.02$\pm$3.213  & 42.64$\pm$1.751 & 46.31$\pm$2.036  & 46.77$\pm$2.362  & 47.26$\pm$1.186
 \\
&    &  & \cellcolor{lightred} ASR  & \cellcolor{lightred}100.00$\pm$0.000 & \cellcolor{lightred}100.00$\pm$0.000& \cellcolor{lightred}74.49$\pm$31.490 & \cellcolor{lightred}64.85$\pm$41.325 & \cellcolor{lightred}100.00$\pm$0.000 & \cellcolor{lightred}100.00$\pm$0.000 & \cellcolor{lightred}100.00$\pm$0.000 & \cellcolor{lightred}100.00$\pm$0.000 & \cellcolor{lightred}99.60$\pm$1.160 & \cellcolor{lightred}100.00$\pm$0.000 & \cellcolor{lightred}100.00$\pm$0.000& \cellcolor{lightred}100.00$\pm$0.000\\
    \hline \hline
\multirow{3}{*}{DM} & \multirow{3}{*}{\rotatebox{90}{CIFAR10}} & \multirow{3}{*}{50} & Baseline& 48.50$\pm$0.470 & 54.19$\pm$0.405 & 20.23$\pm$4.414 & 35.50$\pm$1.040 & 42.50$\pm$0.604 & 42.97$\pm$0.590 & 28.18$\pm$1.521 & 29.88$\pm$1.151 & 25.77$\pm$0.577 & 26.11$\pm$0.881 & 26.36$\pm$0.830 & 27.79$\pm$0.654\\ [0.2ex]
 &    & & BA & 33.42$\pm$0.688
& 35.26$\pm$0.620 & 20.39$\pm$5.763 & 23.99$\pm$5.312 & 29.00$\pm$0.666 & 29.45$\pm$0.695 & 18.35$\pm$0.629 & 20.23$\pm$1.523 & 15.92$\pm$0.691 & 16.04$\pm$0.541 & 17.92$\pm$0.868 & 18.49$\pm$0.726\\ [0.2ex]
&    &  & \cellcolor{lightred} ASR & \cellcolor{lightred}99.91$\pm$0.108 & \cellcolor{lightred}99.25$\pm$0.406 & \cellcolor{lightred}84.05$\pm$30.952 & \cellcolor{lightred}97.89$\pm$3.954 & \cellcolor{lightred}100.00$\pm$0.000 & \cellcolor{lightred}100.00$\pm$0.000 &  \cellcolor{lightred}100.00$\pm$0.000 & \cellcolor{lightred}100.00$\pm$0.000 & \cellcolor{lightred}99.88$\pm$0.306 & \cellcolor{lightred}98.93$\pm$1.866 & \cellcolor{lightred}100.00$\pm$0.000 & \cellcolor{lightred}100.00$\pm$0.000\\ [0.2ex] \hline \hline
\multirow{15}{*}{DSA} & \multirow{6}{*}{\rotatebox{90}{CIFAR10}} & \multirow{3}{*}{1} & Baseline & 26.24$\pm$0.566 & 26.69$\pm$0.807 & 19.25$\pm$1.465 & 17.04$\pm$2.166 & 21.99$\pm$0.714 & 22.23$\pm$1.054 & 13.53$\pm$0.867 & 15.07$\pm$0.752 & 23.49$\pm$1.063 & 25.00$\pm$0.957 & 21.18$\pm$1.138 & 22.06$\pm$1.179 \\
                      &  & & BA    & 24.09$\pm$0.719
& 24.70$\pm$0.552 & 18.65$\pm$1.322
& 16.50$\pm$3.057 & 19.01$\pm$1.579
& 21.60$\pm$1.066 & 12.24$\pm$1.515
& 13.12$\pm$0.840 & 20.89$\pm$0.806
& 22.48$\pm$0.433 & 18.24$\pm$1.034
& 19.38$\pm$1.397      \\ 
                    &  &  & \cellcolor{lightred}ASR     & \cellcolor{lightred}100.00$\pm$0.004
& \cellcolor{lightred}99.99$\pm$0.029 & \cellcolor{lightred}68.90$\pm$34.971
& \cellcolor{lightred}67.72$\pm$33.232 & \cellcolor{lightred}97.38$\pm$5.153
& \cellcolor{lightred}99.76$\pm$0.714 & \cellcolor{lightred}89.64$\pm$29.715
& \cellcolor{lightred}90.52$\pm$26.595 & \cellcolor{lightred}96.27$\pm$6.907
& \cellcolor{lightred}99.83$\pm$0.216 & \cellcolor{lightred}89.81$\pm$18.501
& \cellcolor{lightred}97.34$\pm$5.506       \\ \cline{3-16}
& & \multirow{3}{*}{10} & Baseline & 38.17$\pm$0.600 & 44.23$\pm$0.549 & 14.48$\pm$2.514 & 27.51$\pm$0.830 & 33.25$\pm$0.905 & 37.30$\pm$1.589  &19.60$\pm$2.356 & 24.15$\pm$1.801 & 24.53$\pm$0.715 & 29.56$\pm$0.585 & 21.10$\pm$1.343 & 25.50$\pm$2.116\\
&    & & BA       & 32.60$\pm$0.672
& 34.82$\pm$0.449& 17.92$\pm$5.770
& 22.32$\pm$2.974& 26.54$\pm$1.061
& 31.21$\pm$1.015& 17.80$\pm$1.006
& 19.15$\pm$1.476& 19.62$\pm$1.298
& 23.82$\pm$0.696& 18.70$\pm$0.939
& 20.40$\pm$2.199
   \\ 
&    &  & \cellcolor{lightred} ASR     & \cellcolor{lightred}100.00$\pm$0.000
& \cellcolor{lightred}100.00$\pm$0.000& \cellcolor{lightred}46.18$\pm$43.370
& \cellcolor{lightred}47.10$\pm$33.155& \cellcolor{lightred}100.00$\pm$0.000
& \cellcolor{lightred}100.00$\pm$0.000& \cellcolor{lightred}99.99$\pm$0.042
& \cellcolor{lightred}100.00$\pm$0.000& \cellcolor{lightred}98.22$\pm$5.314
& \cellcolor{lightred}100.00$\pm$0.006& \cellcolor{lightred}99.34$\pm$1.842
& \cellcolor{lightred}100.00$\pm$0.003\\  \cline{2-16} 
 & \multirow{6}{*}{\rotatebox{90}{CIFAR100}} & \multirow{3}{*}{1} & Baseline   & 8.78$\pm$0.402
  & 9.99$\pm$0.461  & 1.24$\pm$0.145
  & 2.55$\pm$0.649  & 6.15$\pm$0.205
  & 8.27$\pm$0.579  & 2.07$\pm$0.263
  & 3.11$\pm$0.686  & 2.94$\pm$0.228
  & 5.63$\pm$0.334  & 3.31$\pm$0.231
  & 5.15$\pm$0.507
 \\
& & & BA & 7.66$\pm$0.262 & 9.10$\pm$0.242 & 2.41$\pm$0.524 & 1.97$\pm$1.109 & 5.79$\pm$0.509 & 7.35$\pm$0.468 & 2.23$\pm$0.234 & 3.05$\pm$0.466 & 2.20$\pm$0.288 & 4.07$\pm$0.523 & 3.14$\pm$0.356 & 4.27$\pm$0.499\\
& & & \cellcolor{lightred}  ASR & \cellcolor{lightred}95.31$\pm$5.524 & \cellcolor{lightred}98.43$\pm$1.799 & \cellcolor{lightred}19.75$\pm$24.239 & \cellcolor{lightred}1.21$\pm$2.263 & \cellcolor{lightred}90.56$\pm$20.012 & \cellcolor{lightred}99.69$\pm$0.595 & \cellcolor{lightred}82.52$\pm$28.957 &\cellcolor{lightred} 98.61$\pm$3.621 & \cellcolor{lightred}16.68$\pm$17.132 & \cellcolor{lightred}79.71$\pm$20.856 & \cellcolor{lightred}84.15$\pm$24.050 & \cellcolor{lightred}82.86$\pm$28.564 \\ \cline{3-16}
 & & \multirow{3}{*}{10} & Baseline & 16.28$\pm$0.365 & 23.28$\pm$0.364 & 5.84$\pm$0.848 & 14.69$\pm$0.601 & 11.79$\pm$0.466 & 17.81$\pm$0.484 & 4.43$\pm$0.418 & 7.17$\pm$0.520 & 6.15$\pm$0.416 & 7.83$\pm$0.476 & 5.58$\pm$0.509 & 7.61$\pm$0.524
 \\ 
 &    & & BA  & 7.31$\pm$0.180
& 8.35$\pm$0.133 & 7.14$\pm$1.610
& 6.89$\pm$2.915& 7.23$\pm$0.214
& 8.23$\pm$0.171& 3.18$\pm$0.415
& 4.87$\pm$0.509& 3.85$\pm$0.375
& 4.96$\pm$0.339& 3.53$\pm$0.441
& 4.83$\pm$0.351
\\ 
&    &  & \cellcolor{lightred} ASR  & \cellcolor{lightred}42.76$\pm$9.156
& \cellcolor{lightred}41.90$\pm$6.128& \cellcolor{lightred}35.68$\pm$7.267
& \cellcolor{lightred}25.76$\pm$27.264& \cellcolor{lightred}96.23$\pm$2.352
& \cellcolor{lightred}60.21$\pm$11.040& \cellcolor{lightred}97.17$\pm$5.463
& \cellcolor{lightred}75.47$\pm$17.501& \cellcolor{lightred}95.90$\pm$2.960
& \cellcolor{lightred}75.79$\pm$13.136& \cellcolor{lightred}98.16$\pm$1.720
& \cellcolor{lightred}67.59$\pm$8.190  \\  \cline{2-16}
 & \multirow{6}{*}{\rotatebox{90}{FashionMNIST}} & \multirow{3}{*}{1} &  Baseline &  67.89$\pm$1.046 & 69.22$\pm$0.821 & 42.81$\pm$4.420 & 45.75$\pm$15.920 & 53.03$\pm$3.175 & 57.14$\pm$2.398 & 21.53$\pm$3.859 & 27.98$\pm$5.190 & 62.66$\pm$1.644 & 66.32$\pm$1.923 & 57.96$\pm$2.942 & 58.88$\pm$1.684\\
 &    & & BA  & 61.32$\pm$1.001 & 62.55$\pm$1.303 & 32.27$\pm$13.479 & 32.55$\pm$15.653 & 47.77$\pm$1.926 & 49.78$\pm$1.856 & 19.57$\pm$3.460 & 28.21$\pm$6.489& 56.31$\pm$0.988
& 57.36$\pm$2.118& 49.29$\pm$3.731
& 50.07$\pm$3.197\\
&    &  & \cellcolor{lightred} ASR   & \cellcolor{lightred}100.00$\pm$0.000
& \cellcolor{lightred}100.00$\pm$0.000& \cellcolor{lightred}83.07$\pm$29.260
& \cellcolor{lightred}67.64$\pm$40.277& \cellcolor{lightred}100.00$\pm$0.000
& \cellcolor{lightred}100.00$\pm$0.000& \cellcolor{lightred}100.00$\pm$0.000
& \cellcolor{lightred}100.00$\pm$0.000& \cellcolor{lightred}100.00$\pm$0.000
& \cellcolor{lightred}100.00$\pm$0.000& \cellcolor{lightred}100.00$\pm$0.000
& \cellcolor{lightred}100.00$\pm$0.000  \\ \cline{3-16}
& &    \multirow{3}{*}{10} & Baseline & 73.23$\pm$0.453 & 77.73$\pm$0.353 & 19.56$\pm$5.134 & 61.82$\pm$1.605 & 67.66$\pm$2.521 & 79.22$\pm$0.585 & 52.64$\pm$3.115 & 58.65$\pm$1.668 & 64.18$\pm$2.348 & 70.63$\pm$1.320 & 65.91$\pm$1.782 & 76.19$\pm$1.384\\
 &    & & BA & 71.69$\pm$0.490
& 73.11$\pm$0.459& 28.53$\pm$15.561
& 41.16$\pm$13.745& 66.85$\pm$2.491
& 69.86$\pm$1.348& 47.12$\pm$2.798
& 55.45$\pm$3.117& 51.91$\pm$1.591
& 60.03$\pm$1.326& 48.86$\pm$2.602
& 62.15$\pm$3.643 
 \\
&    &  & \cellcolor{lightred} ASR  & \cellcolor{lightred}100.00$\pm$0.000
& \cellcolor{lightred}100.00$\pm$0.000 & \cellcolor{lightred}79.51$\pm$30.277
& \cellcolor{lightred}80.63$\pm$34.301 & \cellcolor{lightred}100.00$\pm$0.000
& \cellcolor{lightred}100.00$\pm$0.000& \cellcolor{lightred}100.00$\pm$0.000
& \cellcolor{lightred}100.00$\pm$0.000& \cellcolor{lightred}100.00$\pm$0.000
& \cellcolor{lightred}100.00$\pm$0.000& \cellcolor{lightred}100.00$\pm$0.000
& \cellcolor{lightred}100.00$\pm$0.000\\
    \hline \hline \multirow{6}{*}{MTT} & \multirow{6}{*}{\rotatebox{90}{CIFAR10}} & \multirow{3}{*}{1} & Baseline   & 38.76$\pm$1.091
  & 38.80$\pm$1.235  & 10.87$\pm$1.084
  & 14.16$\pm$3.897  & 17.97$\pm$1.501
  & 21.49$\pm$1.605  & 11.37$\pm$1.230
  & 11.04$\pm$0.639  & 14.02$\pm$0.912
  & 14.79$\pm$0.698  & 16.76$\pm$0.992
  & 16.99$\pm$1.019
   \\
                      &  & & BA              & 31.87$\pm$1.087
& 32.58$\pm$0.841 & 11.27$\pm$2.660
& 12.27$\pm$3.573 & 18.11$\pm$1.473
& 19.73$\pm$1.180 & 10.62$\pm$0.641
& 11.52$\pm$0.930 & 12.29$\pm$1.076
& 13.82$\pm$1.125 & 16.28$\pm$1.209
& 16.45$\pm$0.811\\ 
                    &  &  & \cellcolor{lightred}ASR     & \cellcolor{lightred}100.00$\pm$0.000
& \cellcolor{lightred}100.00$\pm$0.000 & \cellcolor{lightred}70.29$\pm$45.392
& \cellcolor{lightred}72.69$\pm$37.554 & \cellcolor{lightred}100.00$\pm$0.000
& \cellcolor{lightred}100.00$\pm$0.000 & \cellcolor{lightred}100.00$\pm$0.000
& \cellcolor{lightred}98.89$\pm$2.271 & \cellcolor{lightred}59.24$\pm$36.482
& \cellcolor{lightred}92.44$\pm$10.594 & \cellcolor{lightred}99.94$\pm$0.123
&\cellcolor{lightred} 99.64$\pm$1.068                     \\ \cline{3-16}
& & \multirow{3}{*}{10} & Baseline   & 43.81$\pm$0.709
  & 51.65$\pm$0.926  & 16.82$\pm$3.091
  & 26.73$\pm$1.160  & 33.44$\pm$0.593
  & 34.35$\pm$1.110  & 23.33$\pm$2.895
  & 25.66$\pm$1.848  & 15.58$\pm$0.367
  & 16.17$\pm$0.829  & 19.53$\pm$1.148
  & 21.22$\pm$0.675\\
&    & & BA         & 34.81$\pm$0.436
& 37.81$\pm$0.569 & 24.13$\pm$4.853
& 11.66$\pm$2.302 & 25.00$\pm$1.176
& 25.92$\pm$1.154 & 17.01$\pm$1.617
& 20.44$\pm$1.647 & 12.33$\pm$0.309
& 12.74$\pm$0.672 & 14.73$\pm$0.831
& 16.06$\pm$1.442 
   \\ 
&    &  & \cellcolor{lightred} ASR    & \cellcolor{lightred}100.00$\pm$0.000
& \cellcolor{lightred}100.00$\pm$0.000 & \cellcolor{lightred}94.08$\pm$5.960
& \cellcolor{lightred}54.31$\pm$43.090 & \cellcolor{lightred}100.00$\pm$0.000
& \cellcolor{lightred}100.00$\pm$0.000 & \cellcolor{lightred}100.00$\pm$0.000
& \cellcolor{lightred}100.00$\pm$0.000 & \cellcolor{lightred}79.19$\pm$13.521
& \cellcolor{lightred}71.56$\pm$31.169 & \cellcolor{lightred}100.00$\pm$0.000
&\cellcolor{lightred} 100.00$\pm$0.000 \\  
    \hline
\end{tabular}
}
\vspace{-10pt}
\end{table*}

\subsection{Experiments about Different Training Strategies}
In this subsection, we evaluate the impact of different training strategies on the effectiveness of our attack based on DC. Specifically, we analyze performance across different numbers of training epochs in user-side training, and we treat the performance of models trained directly on the benign distilled dataset as the baseline. As shown in Table~\ref{tab:stra}, our attack remains highly effective across different numbers of training epochs, consistently maintaining a high ASR while inducing minimal BA degradation. Besides, our method demonstrates strong generalizability across various raw datasets and different images per class (IPC) settings, ensuring its robustness in diverse scenarios.

In previous experiments, we assume that the user-side model was identical to the attacker's model. To further validate the robustness of our method, we investigate a more challenging scenario where the user-side model differs from the attacker's model. We analyze the attack performance under different training strategies, and the results are presented in Figure~\ref{fig:performance}. In this experiment, we use CIFAR-10 as the raw dataset based on DC with setting IPC to~1. As shown in the results, our method remains highly effective, consistently delivering strong attack performance even when the user-side model differs from the attacker's model.

\subsection{Experiment with Different Dataset Distillation Methods}
To further validate the effectiveness of our attack, we extend our experiments to different DD methods, with the results summarized in Table~\ref{tab:tab2}. We conduct evaluations using user-side training strategies of 50 and 100 epochs. As shown in Table~\ref{tab:tab2}, our attack consistently demonstrates strong performance across various DD methods. In most cases, the attack achieves nearly 100\% ASR, effectively embedding the backdoor into the distilled dataset, regardless of the specific DD approach employed. Additionally, the BA degradation remains within an acceptable range, which indicates that the overall utility of the dataset is well preserved. These results confirm the generalizability and robustness of our proposed attack method, demonstrating its effectiveness across different distillation strategies while maintaining the performance of downstream tasks.

\subsection{Visualization}
\begin{figure}[!t]
  \centering
  \includegraphics[width=\linewidth]{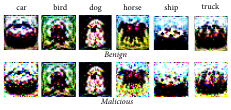}
  \vspace{-20pt}
\caption{Visualization of benign and malicious distilled data. Without a direct comparison, users may struggle to sense the subtle differences due to the inherent abstraction of distilled datasets.}
  \label{fig:visua}
\end{figure}

Figure~\ref{fig:visua} presents a visual comparison between benign and malicious distilled datasets. The first row displays examples of benign distilled images, while the second row illustrates their malicious counterparts after backdoor injection. Due to the inherent abstraction of distilled datasets, these images inherently lack fine-grained details, making it challenging for users to discern their authenticity based on individual distilled samples. This abstraction further facilitates the attack, as the malicious modifications remain visually subtle and difficult to detect. Despite these seemingly minor perturbations, the backdoor triggers embedded in the malicious dataset remain highly effective, ensuring that models trained on this data reliably respond to the attacker's intended inputs.

\begin{table}[!t]
    \centering
        \caption{Attack performance with different attacker-side downstream models (Avg $\pm$ STD, \%).}
    \resizebox{\columnwidth}{!}{
    \begin{tabular}{c|c|c|c|c}
    \hline
 Model & & \multicolumn{1}{c}{DC}  & \multicolumn{1}{|c}{DSA} & \multicolumn{1}{|c}{MTT}  \\ \cline{1-5}
\multirow{3}{*}{AlexNet} & Baseline   & 21.48$\pm$0.729 & 27.51$\pm$0.830 &  26.73$\pm$1.160 \\
& BA   & 16.56$\pm$6.264 & 21.65$\pm$1.919 & 19.86$\pm$3.226 \\
& ASR   &23.04$\pm$38.896  & 21.36$\pm$16.998 & 18.33$\pm$18.708 \\ \hline
\multirow{3}{*}{VGG11} & Baseline   & 25.66$\pm$0.952 & 35.27$\pm$0.497 & 34.35$\pm$1.110\\
& BA   &  23.58$\pm$0.991 & 29.01$\pm$1.107& 20.54$\pm$1.054
\\
& ASR  & 68.89$\pm$20.244 & 49.77$\pm$26.597 & 77.21$\pm$20.196
\\ \hline
\multirow{3}{*}{VGG16} & Baseline   & 26.16$\pm$1.630 & 24.15$\pm$1.801 & 25.66$\pm$1.848  \\
& BA &  21.01$\pm$2.285& 21.69$\pm$1.418
 & 21.57$\pm$1.493 
\\
& ASR &  56.70$\pm$37.613 & 17.99$\pm$24.918
 & 73.78$\pm$35.649\\ \hline
    \end{tabular}
    }
    \vspace{-5pt}
    \label{tab:diff_mal_model}
\end{table}

\subsection{Ablation Study}

In previous experiments, we used ConvNet as the attacker's downstream model. In this experiment, we evaluate the impact of different model architectures on the effectiveness of our attack. To demonstrate the generalizability of our method, we conduct experiments on CIFAR-10 distilled using different DD methods, with IPC set to 10. The results, presented in Table~\ref{tab:diff_mal_model}, indicate that our attack remains highly effective across various model architectures. It can be seen that our threat model operates under relatively weak assumptions, making it highly practical in real-world scenarios. Despite these relaxed constraints, our attack maintains strong performance across various settings.

\begin{figure}[!t]
  \centering
  \includegraphics[width=.95\linewidth]{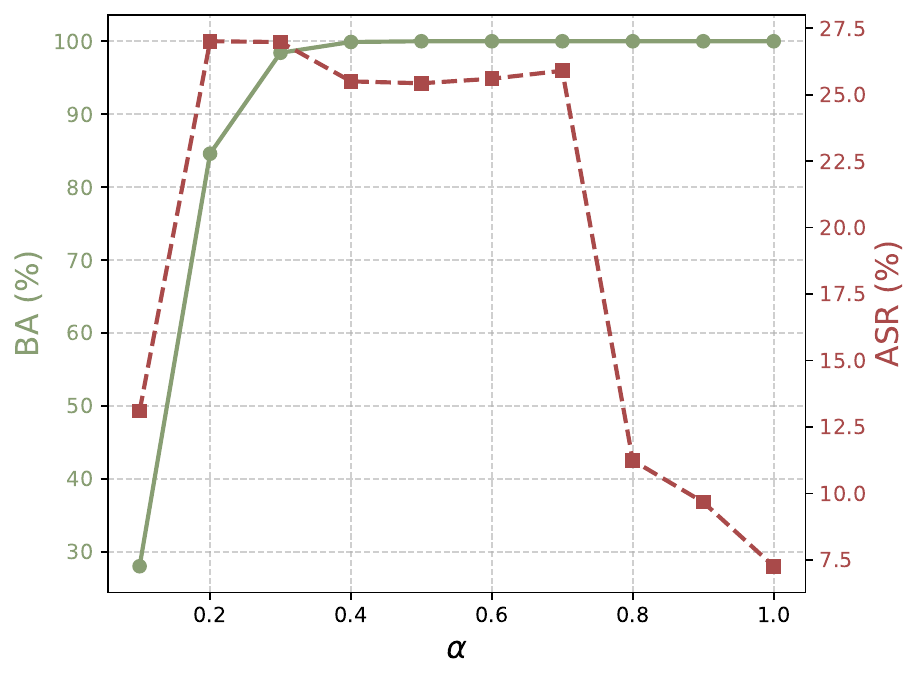}
  \vspace{-10pt}
  \caption{The performance under different $\alpha$ in Eq.~\eqref{eq:loss}.}
  \vspace{-5pt}
  \label{fig:alpha}
\end{figure}

To further analyze the impact of different components in our method, we conduct an ablation study on the effect of $\alpha$ in Eq.~\eqref{eq:loss}, which balances the tradeoff between attack effectiveness and benign task performance. In this experiment, we use the CIFAR-10 dataset distilled by the DC method, with IPC set to 1. The results are shown in Figure~\ref{fig:alpha}. As $\alpha$ increases, ASR remains high, but the performance degradation on benign tasks becomes more pronounced. This occurs because a larger $\alpha$ emphasizes backdoor retention, potentially sacrificing the utility of the distilled dataset. To achieve an optimal balance between attack effectiveness and performance retention, we set $\alpha$ to 0.5 in our main experiments.

\subsection{Computational Complexity}
Our attack method is highly efficient and lightweight. To evaluate its computational cost, we conduct experiments on a CIFAR-10 dataset distilled by the DC method with IPC set to 1. The computational complexity varies based on the number of reconstructed conceptual archetypes, and the results are summarized in Table~\ref{tab:compu}. It can be seen that different numbers of conceptual archetypes achieve effective attacks while maintaining minimal impact on benign performance. Therefore, we recommend using $m = 5$ as the default setting, as it provides a balance between efficiency and attack effectiveness.

The total attack time consists of two parts: conceptual archetype reconstruction and malicious distilled dataset synthesis. In the first phase, reconstructing each conceptual archetype requires only 0.53s, and under our default setting of five archetypes per class, this step takes approximately 26.5s for CIFAR-10. In the second phase, due to the small size of the distilled dataset, each epoch of synthesizing the malicious distilled dataset takes only 1.5 seconds on a single NVIDIA GTX 3090. Consequently, the entire attack can be completed in less than one minute. This minimal time overhead makes the attack virtually imperceptible to users, as they are unlikely to notice any delays that could suggest an ongoing attack.


\begin{table}[!t]
    \centering
        \caption{Computational complexity and attack performance with varying numbers of conceptual archetypes.}
        \vspace{-5pt}
        \label{tab:compu}
\resizebox{\columnwidth}{!}{
\begin{tabular}{c|cc|ccc}
\hline
\multirow{2}{*}{} & \multicolumn{2}{c|}{Performance (\%)} & \multicolumn{3}{c}{Time (s)} \\ \cline{2-6} 
                  & \multicolumn{1}{c}{BA}                & ASR               & Per Image  & Per Epoch & All \\ \cline{1-6} 
$m=5$                 &   25.79      &    100.00               &  0.53          &   1.50        &   $0.53\times 5 \times 10 + 1.50\times 10 = 41.5$ \\ \hline
$m=10$                 &  25.59                 & 100.00                  &       0.53       &     1.65      & $0.53\times 10 \times 10  + 1.65\times 10 = 69.50$   \\ \hline
$m=20$                 &     25.36              &      100.00             &     0.53         &     1.96      &  $0.53\times 20 \times 10  + 1.96\times 10 = 125.60$   \\ \hline
\end{tabular}
}
\vspace{-15pt}
\end{table}

\section{Conclusion}
In this paper, we propose a novel backdoor attack method targeting distilled datasets, which enables successful backdoor injection without requiring access to raw data, knowledge of the DD process, or modifications to the data owner's pipeline. Our approach leverages the intrinsic properties of DD by reconstructing conceptual archetypes that align with the latent representations of real images, thereby bridging the gap between distilled and real data. We then embed backdoor information into the distilled dataset to ensure a consistent optimization trajectory with benign training, effectively concealing malicious behavior. Extensive experiments across various DD methods, raw datasets, training strategies, and downstream architectures demonstrate the effectiveness, generalizability, and stealthiness of our method. Our findings reveal a critical security vulnerability in dataset distillation, challenging the common belief that distilled datasets are inherently resistant to backdoor attacks~\cite{liu2023backdoor}. We hope this work raises awareness of potential threats and encourages further research into defense mechanisms that ensure the security and trustworthiness of distilled datasets. In future work, we will develop defense algorithms to mitigate the proposed attack.


\bibliographystyle{ACM-Reference-Format}
\bibliography{sample-base}










\end{document}